\documentstyle[11pt,aaspp4]{article}     

\received{May 20, 1997}
\accepted{August 27, 1997}

\journalid{000}{000}
\articleid{00}{00}

\lefthead{Courteau}
\righthead{Optical Rotation Curves of Spirals}
\slugcomment{\it Accepted for publication in the Astronomical Journal}

\begin{document}


\def\nowidth#1{\hidewidth#1\hidewidth}

\def\gtsima{$\, \buildrel > \over \sim \,$}
\def\ltsima{$\, \buildrel < \over \sim \,$}
\def\simgt{\lower.5ex\hbox{\gtsima}}
\def\simlt{\lower.5ex\hbox{\ltsima}}
\def\gtaprx{\mathrel{\vcenter{\offinterlineskip \hbox{$>$}
    \kern 0.3ex \hbox{$\sim$}}}}
\def\ltaprx{\mathrel{\vcenter{\offinterlineskip \hbox{$<$}
    \kern 0.3ex \hbox{$\sim$}}}}
\def \yskip{\penalty-50\vskip3pt plus3pt minus2pt}
\def \yyskip{\penalty-100\vskip6pt plus6pt minus4pt}

\def \er#1       {\hbox{\rm {#1}}}
\def \eb#1       {\hbox{\bf {#1}}}

\def \v22{V_{2.2}}
\def\kms{\ifmmode {\rm \, km \, s^{-1}} \else $\rm \,km \, s^{-1}$\fi}  
\def\ksmpc{\ifmmode{\rm km}\,{\rm s}^{-1}\,{\rm Mpc}^{-1}\else km$\,$s$^{-1}\,$Mpc$^{-1}$\fi}
\def   \etal {{\it et al.~}}
\def   \eg   {{\it e.g.~}}
\def   \ie   {{\it i.e.~}}
\def   \vs   {{\it vs~}}
\def   \cf   {{cf.~}}
\def   \etc  {{\it etc.}}
\def   \apriori  {{\em a priori}}
\def   \aposteriori{{\em a posteriori}}
\def   \ibid     {{\it ibid.}}
\def   \opcit    {{\it opcit.}}
\def   \vis      {{\it vis}}
\def   \dnsigma  {$D_n$-$\sigma$}
\def   \lsol     {$L_\odot$}
\def   \msol{\ifmmode M_\odot\else$M_\odot$\fi}
\def   \onesigma{$1$-$\sigma$}
\def   \dline{\noalign{\hrule height1pt\vskip 2pt\hrule height 1pt\vskip 4pt}}
\def \bline{\noalign{\vskip 5pt\hrule height1pt}}
\def \pp{\par\yskip\noindent\hangindent 0.4in \hangafter 1}
\def \reference#1#2#3#4 {\pp#1, {\it#2}, {\bf#3}, #4.}
\def \refpp{\par\yskip\noindent\hangindent 0.74in \hangafter 1}
\def \subp#1{\hangindent=6em \hangafter 1 #1 }
\def \littleprime{\ifmmode{\scriptscriptstyle \prime }
     \else{\hbox{$\scriptscriptstyle \prime$ }}\fi}
\def \littless{\ifmmode{\scriptscriptstyle s }
     \else{\hbox{$\scriptscriptstyle s $ }}\fi}
\def \littlemm{\ifmmode{\scriptscriptstyle m }
     \else{\hbox{$\scriptscriptstyle m $ }}\fi}
\def \littlecirc{\ifmmode{\scriptscriptstyle \circ }
     \else{\hbox{$\scriptscriptstyle \circ $ }}\fi}
\def \littlehour{\ifmmode{\scriptscriptstyle h }
     \else{\hbox{$\scriptscriptstyle m $}}\fi}
\def \toph{\raise .9ex \hbox{\littlehour}}
\def \hourpoint{\hbox to 2pt{}\rlap{\hskip -.5ex \toph}.\hbox to 2pt{}}
\def \topemm{\raise .9ex \hbox{\littlemm}}
\def \magpoint{\hbox to 2pt{}\rlap{\hskip -.5ex \topemm}.\hbox to 2pt{}}
\def \arcss{\raise .9ex \hbox{\littless}}
\def \arcsspoint{\hbox to 1pt{}\rlap{\arcss}.\hbox to 2pt{}}
\def \sdeg{\raise .9ex \hbox{\littlecirc}}
\def \degpoint{\hbox to 1pt{}\rlap{\sdeg}.\hbox to 2pt{}}
\def \arcmin{\raise .9ex \hbox{\littleprime}}
\def \arcminpoint{\hbox to 1pt{}\rlap{\arcmin}.\hbox to 2pt{}}
\def \arcsec{\raise .9ex \hbox{\littleprime\hskip-3pt\littleprime}}
\def \arcsecpoint{\hbox to 1pt{}\rlap{\arcsec}.\hbox to 2pt{}}
\def \ra#1#2#3{{#1}^{\er{\sevenrm h}}{#2}^{\er{\sevenrm m}}
     {#3}^{\er{\sevenrm s}}}
\def \dec#1#2#3{{#1}^{\circ}{#2}'{#3}''}
\def\deg{$^\circ$}
\def\degc{^\circ \er{C}}
\def\sol{$_\odot$}
\def\sun{$_{\scriptscriptstyle\odot}$}
\def\Msun{\ifmmode \,{\rm M}_{\odot}        
          \fi}
\def\Lsun{\ifmmode \,{\rm L}_{\odot}        
          \else $\,${\rm L}$_{\odot}$
          \fi}
\def\Rsun{\ifmmode \,{\rm R}_{\odot}        
          \else $\,${\rm R}$_{\odot}$
          \fi}
\def \msun     {\Msun}
\def \stel     {_\ast}
\def \earth    {_\oplus}
\def \losol#1  {({\hbox{#1}}/{\hbox{#1}\sol})}
\def \bosol#1  {\left({{\hbox{#1}}\over{\hbox{#1}\sol}}\right)}
\def \eff      {$\si{eff}$}
\def \hub      {$H_{\hbox{\scriptsize 0}}$}
\def \hunit    {\kms~Mpc$^{\hbox{\scriptsize -1}}$}
\def \dv       {$\Delta V$}
\def \dvs      {$\Delta V$'s}
\def \chisqr   {$\chi^2$}
\def \chidof   {$\chi^2_\nu$}
\def \ion#1#2{#1$\;${\small\rm{#2}}\relax}
\def \halpha   {{H$\alpha$\ }}
\def \ha       {H$\alpha$\ }
\def \hbeta    {H$\beta$ }
\def \hgamma   {H$\gamma$ }
\def \caii     {\ion{Ca}{II}}
\def \civ      {\ion{Ca}{IV}}
\def \fei      {\ion{Fe}{I}}
\def \feii     {\ion{Fe}{II}}
\def \hi       {\ion{H}{I} }
\def \hii      {\ion{H}{II} }
\def \hei      {\ion{He}{I}}
\def \heii     {\ion{He}{II}}
\def \mgi      {\ion{Mg}{I}}
\def \nad      {\ion{Na}{I}~D}
\def \ni       {\ion{N}{I} }
\def \nii      {[\ion{N}{II}] }
\def \oi       {[\ion{O}{I}] }
\def \oii      {[\ion{O}{II}] }
\def \oiii     {[\ion{O}{III}] }
\def \sii      {[\ion{S}{II}] }
\def \siii     {[\ion{S}{III}] }
\def \nei      {\ion{Ne}{I} }
\def \neiii    {[\ion{Ne}{III}] }
\def \neiv     {[\ion{Ne}{IV}] }
\def \nev      {[\ion{Ne}{V}] }

\title{Optical Rotation Curves and Linewidths for Tully-Fisher 
       Applications \altaffilmark{1}}

\author{St\'ephane Courteau}
\affil{Kitt Peak National Observatory, 950 North Cherry Avenue, Tucson, AZ 85726
\altaffilmark{2}}

\altaffiltext{1}{Based on observations made at Lick (UCO) and Las Campanas (OCIW) 
  Observatories.}

\altaffiltext{2}{Current Address: Herzberg Institute for Astrophysics, 
 Dominion Astrophysical Observatory, 5071 W. Saanich Rd., Victoria, B.C. V8X 4M6. \
 email: stephane.courteau@hia.nrc.ca}

\begin{abstract}

We present optical long-slit rotation curves (RCs) for 304 northern Sb-Sc UGC galaxies
originally selected for Tully-Fisher (TF) applications.  Matching $r$-band photometry 
exists for each galaxy.  We describe the procedures of RC extraction 
and construction of optical profiles analogous to 21cm integrated 
linewidths.  More than 20\% of the galaxies were observed twice or more,
allowing for a proper determination of systematic errors.
Various measures of maximum rotational velocity to be used as input
in the TF relation are tested on the basis of their repeatability,
minimization of TF scatter, and match with 21cm linewidths.   
The best measure of TF velocity, $\v22$, is given at the location 
of peak rotational velocity of a pure exponential disk. 
An alternative measure to $\v22$ which makes no assumption about the luminosity
profile or shape of the rotation curve is $V_{hist}$, the 20\% width of the 
velocity histogram, though the match with 21cm linewidths is not as good.
We show that optical TF calibrations yield internal scatter comparable to, 
if not smaller than, the best calibrations based on single-dish 21cm radio 
linewidths.  Even though resolved \hi RCs are more extended than 
their optical counterpart, a tight match between optical and radio linewidths 
exists since the bulk of the \hi surface density is enclosed within the optical 
radius. 

We model the 304 RCs presented here and a sample of 958 curves from 
Mathewson \etal (1992) with various fitting functions.  
An arctan function 
provides an adequate simple fit (not accounting for non-circular motions 
and spiral arms).  More elaborate, empirical models
may yield a better match at the expense of strong covariances. 
We caution against physical or ``universal'' parameterizations 
for TF applications. 

\end{abstract}

\section{Introduction}

Since the pioneering work of Oepik (1922), 
it has been 
understood that rotational velocities and apparent magnitudes could be used 
to infer extragalactic distances (\cf \eg Rubin 1995).  
Work on correlations between rotational velocities and
diameter or luminosity in the 1960s and early 1970s (Roberts 1969, 
Bottinelli \etal 1971, Balkowski \etal 1974, Shostak 1975, Roberts 1975, 
Sandage \& Tammann 1976) led to the definitive formulation 
of the Tully-Fisher relation (Tully \& Fisher 1977; hereafter TF) which has
become one of the most powerful and widely used tools for studying cosmic velocity 
fields in the local universe (cf. Strauss \& Willick 1995 for a review). 

The TF relation is an empirical correlation between some suitably defined 
measure of the ``peak'' circular velocity (or linewidth), $V$, and the
total absolute magnitude, $M$, of a spiral galaxy, such that $M=a\log V + b$.  
Together, these two parameters
define a plane of structural parameters for spiral galaxies which may reflect
the way they were initially formed or perhaps suggest the presence of 
self-regulating processes of star formation in galactic disks. 
There are several models of galaxy formation which predict 
of a correlation between the luminosity and rotation velocity for spiral
galaxies (Fall \& Efstathiou 1980, Faber 1982, Gunn 1982, Blumenthal \etal\ 1984,  
Ashman 1990, Cole \etal 1994, Eisenstein \& Loeb 1995, Eisenstein 1996). A correlation 
is expected, as both luminosity and disk rotational velocity depend on galaxy 
mass.  An alternative model for the TF relation involves dissipative galaxy 
formation in which regulated star formation, or the ability of the galaxy 
to retain gas ejected from stars and supernovae, is determined by the amount 
of baryonic matter in the halo.  In this model, the TF relation arises as 
a natural ``prediction'' with negligible scatter (White 1991, Kauffmann \etal 1993, 
Silk 1995, 1997).  The residuals in the TF relation must therefore reflect fluctuations in the 
detailed formation and enrichment history of galactic halos.  Evolution and 
environmental effects are therefore likely to modify the shape of the TF 
relation in different environments and would suggest that a true universal TF 
relation $L \propto V^3$ does not exist. 

Various TF calibrations will be obtained depending upon the choice 
of photometric bandpass and measure of rotational speed.
These, in addition to corrections for various intrinsic, environmental,
and instrumental biases, all contribute to the shape, 
zero-point, and scatter of the TF relation. 
It is now well understood that the slope and zero-point of the TF relation 
depend strongly on aperture and bandpass and that a smaller dispersion is 
obtained for near-IR (NIR) bandpasses ($R,I,$ and $H$) than for the $B$ 
or $V$ bands.  As first argued by Aaronson and coworkers
in the 80's (\eg Aaronson \etal 1989 and references therein), infrared photometry 
suffers less from 
Galactic and internal absorption corrections than do bluer bandpasses.
Moreover, it is mostly sensitive to the underlying evolved stellar population 
which best traces the optical mass and thus correlates more tightly with the maximum
rotational speed or, total galaxy mass.  Inclinations are also better determined at 
infrared wavebands since the overall image is smoother (Peletier \& Willner 1993). 

There is a well-known trend of increasing TF slopes toward longer wavelengths; 
representative ranges for the $B, R, I,$ and $H$ bandpasses are roughly 5, 6, 7
and 8\footnote{
Note that this infrared TF slope (Bernstein \etal 1994) is appreciably closer to optical 
slopes than the canonical Aaronson value of $\sim 10$.  The larger slope observed at 
$H$-band by Aaronson \etal (1982) was shown to be largely due to aperture effects 
(Willick 1991). Pierce \& Tully (1992) also obtain a steep slope at H which is mostly
due to their ``inverse'' fitting approach.} respectively (Strauss \& Willick 1995,
Willick \etal 1997, hereafter W97).
The typical slope for a linear TF relation is close to 7.5 for the red
and infrared bands indicating that $L_{\rm R/IR} \propto V_{\rm max}^{3}$. 
The typical intrinsic rms dispersion of the best forward\footnote{\cf Strauss \& Willick (1995)
or Willick \etal (1996) for a discussion of forward vs inverse fitting techniques
and inherent biases.  In this paper, we concern ourselves only with the 
forward approach.} TF calibrations, with carefully 
measured CCD magnitudes and inclinations, is between 0.25-0.40 mag ($1\sigma$) or 
a 12-20\% distance error.  TF scatters as low as 0.1 mag have been reported 
for pruned samples with very specific selection criteria (Bothun \& Mould 1987,
Pierce \& Tully 1992, Rood \& Williams 1993, Bernstein \etal 1994, Raychaudhury \etal 1997, 
hereafter R97) but these do not reflect the normal population sampled in larger TF surveys.  
Recent NIR TF investigations in $r$ and $I$ bandpasses have yielded comparable, 
if not smaller, TF dispersions 
than similar $H$-band TF calibrations (Pierce \& Tully 1992, Willick 1991, 
Han 1991, Courteau 1992 (hereafter C92), Mathewson \etal 1992 (hereafter MAT), 
Peletier \& Willner 1993, Schommer \etal 1993, 
Bernstein \etal 1994, W97, Giovanelli \etal 1997 (hereafter G97), R97). 
No advantage is gained by imaging at yet longer wavelengths. 
As underscored by Rhee (1996), the $K$-band TF scatter is significantly larger 
on account of a less stable sky background and greater sensitivity to stellar 
population effects (\cf Regan \& Vogel 1994, Rhoads 1997). 
The latter will induce greater uncertainty in inclination estimates as well.
TF scatter increases significantly at longer wavelengths where emission is 
mostly dominated by dust.  In fact, the TF relation disappears beyond 10$\mu$m,
though the greater scatter reported in the $L$-band is also largely due to 
higher instrumental noise and smaller galaxy fluxes (Malhotra \etal 1997). 

While photometric bandpass is an important contributor to the TF slope and 
zero-point, the specific definition of linewidths is equally fundamental. 
For example, the TF calibrations for MAT and Han-Mould galaxies are both based 
on I-band imaging yet they differ in slope by as much as a unit due to 
different linewidth definitions (W97).  The former study uses resolved 
optical RCs (ORCs) and the latter is based on integrated radio linewidths. 
There is thus strong impetus to understand systematic effects due to the 
choice of linewidths.  
Besides {\it intrinsic} or {\it cosmic} scatter, raw linewidths {\it per se} are 
the leading source of dispersion in the overall TF error budget 
(Schommer \etal 1993, C92).  Photometric inclination errors are a close second 
(Pierce \& Tully 1988, Peletier \& Willner 1991, Willick 1991, C92).  
The magnitude dispersion induced by inclination scatter 
(for a given internal extinction correction) is only 
$\sim$ 0.04-0.05 mag, and the total error on fully
processed magnitudes is typically 0.06-0.08 mag (Schommer \etal 1993, Courteau 1996).
Linewidths are more severely impacted.  Inclination errors account for about 
0.10 mag of the total linewidth error which is typically 0.20 mag
(C92, G97). 
This is greater than the dispersion of $\sim 0.15$ induced by axisymmetric 
deviations in galactic disks (Franx \& de Zeeuw 1992, Rix \& Zaritzky 1995).  
Other sources collectively labelled as ``cosmic scatter'' may contribute a 
substantial share of the TF error budget but none independently as large as that 
from raw linewidth measurements (W97 show that the overall cosmic scatter is $\sim$ 
0.30 mag).  
One must therefore proceed with Procrustean care when defining a TF linewidth.

It has also been demonstrated that the TF residuals are mostly Gaussian (W97)
and not a function of any likely ``third parameter'' such as central and mean 
surface brightnesses, bulge-to-disk ratios, RC shape, concentration indices, 
morphological type, 
colors, etc. (Schlegel 1995, Courteau \& Rix 1997 (CR97), W97; see however 
G97, Giraud 1986, Bothun \& Mould 1987).  Spiral galaxies, in this sense, truly 
define a one-parameter sequence.  
After correction for various observational effects, the TF dispersion appears 
to have been reduced to its minimum dispersion in the 
luminosity-linewidth plane of spiral galaxies.  

Measurements of TF rotational velocities have primarily relied on single-dish 
radio observations of neutral hydrogen at 21cm.  
However, the restricted sky coverage of the Arecibo and Nan\c cay radio-telescopes, 
lack of correlator power of array instruments, and lower sensitivity of steerable dishes 
constitute crucial limitations for any \hi-based TF survey.  Emission lines traced in the 
optical domain provide a most adequate substitute to 21cm emission.  
The good match between radio and optical velocity widths has been emphasized 
by Thonnard (1983) and further verified by Courteau \& Faber (1988), C92, MAT, 
and Vogt (1994). 
Considerable effort has been invested in recent years in large TF surveys based on 
ORCs or integrated optical profiles (Dressler \& Faber 1990, 
C92, MAT, Schommer \etal 1993, Vogt 1994, Schlegel 1995, G97.)
The goal of this paper is to address technical issues which pertain to 
the extraction and analysis of long-slit optical spectra for TF
applications.

We present below a detailed account of the acquisition and calibration 
of long-slit ORCs for TF applications.  We use a sample of 304 UGC field 
Sb-Sc spirals with \ha spectra and $r$-band photometry
drawn from the collection of field Sb-Sc spirals of Courteau (1992; hereafter C92). 
This sample is also referred to as ``Courteau-Faber'' (hereafter CF). 
Multiple observations for 62 galaxies combine for a total of 132 spectra which
will be used to study systematics and errors.

Sample selection, catalog information, redshift histograms, sky distribution,
and photometric reductions were all presented in Courteau (1996; hereafter Paper I).
The division of this paper is as follow:
In \S 2 we discuss the spectroscopic observations and data reduction. 
The extraction of rotation curves from resolved emission lines is discussed
in \S 3 and in \S 4, we compare our results with the long-slit RCs of Rubin and 
coworkers, Vogt (1994), and the Fabry-Perot RCs of Amram \etal (1992, 1994). 
The reader familiar with RC extractions can skip to \S 5 without loss of continuity. 
We address various issues of RC shapes and fitting models in \S 5,
which includes a brief discussion of the applicability of the 
``Universal Rotation Curve'' of Persic, Salucci, \& Stel (1996; hereafter PSS). 
We introduce various definitions of velocity measurements for TF use both 
from resolved RCs and integrated profiles in \S6. 
In \S 7, we make use of repeat observations to investigate
the internal stability of our data.  Corrections to 
resolved and integrated linewidths are shown in \S 8, and we assemble a
table of corrected data in \S 9.  Optical TF calibrations are derived in
\S 10 and we examine the tight match between optical and radio 
linewidths in \S 11.
Three appendices include A) a presentation of all the CF
RCs, two-sided and folded, and best-fitting models; B) model parameters 
for all fitted CF ORCs and the 958 ORCs from the compilation of MAT; and C) 
a discussion on intensity-weighted centroids and the use of parabolic 
interpolation for binned data.  

Throughout the paper, we make the distinction between three measures of
rotational linewidths.  We shall refer to the velocity separation between the 
approaching and receding sides of the resolved RC as \dv (\eg \dv$_{max}$).  The 
width of any one-dimensional (integrated) rotation profile, be it a fully 
convolved 21cm spectrum or a co-added or unresolved optical spectrum, will be 
called $W$ (\eg $W_{20}, W_{50}$).  A velocity measured from a one-sided
resolved rotation curve will simply be denoted as $V$ (\eg $\v22, V_{iso}, 
V_{max}, V_{opt}$).   Physical scales are calculated using \hub = 70 \hunit. 

\section{Spectroscopic Observations and Data Reduction}

Spectroscopic observations were made at Lick Observatory using the UV Schmidt Cassegrain
Spectrograph on the Shane 3.0m telescope (Miller \& Stone 1987) and at Las Campanas 
with the Modular Spectrograph on the 2.5m du~Pont telescope. 
Table 1 summarizes the main characteristics of both instrumental setups. 
A total of 453 \ha spectra, including repeats,
was collected at Lick (43 nights) and Las Campanas (1 night courtesy of Alan 
Dressler).  Only ten spectra showed no measurable signal. 
The final collection of RCs includes measurements for 304 galaxies, 
62 of them were observed more than once.  Eight galaxies were observed twice 
as a consistency check between Lick and Las Campanas.  Minor axis RCs for 
6 galaxies were also observed to verify the lack of significant velocity gradients. 

\begin{table}[t]
\centering
\begin{tabular}{l l l}
\multicolumn{3}{c}{TABLE 1} \\
\multicolumn{3}{c}{CCD/Telescope Parameters} \\ \hline\hline
& Lick & Las Campanas \\ \hline 
 Telescope & Shane 3.0m & du Pont 2.5m \\  [1pt]
 Spectrograph & UV Schmidt & Modular \\ [1pt]
 Grating & 1200~$\ell$/mm & 1200~$\ell$/mm \\ [1pt] 
 Blaze ($1^{\scriptsize{st}}$ order)& 5000 \AA\ & 7500 \AA\  \\  [1pt]
 Camera (Nikkor) & 58mm f/1.2  & 85mm f/1.4 \\   [1pt]
 Spectral coverage & 6150$-$6950 \AA\ & 6080$-$7290 \AA\ \\   [1pt]
 Slit width & 2$\arcsecpoint$8 $\times$ 2$\arcmin$ & 2$\arcsecpoint$5 $\times$ 2$\arcmin$ \\ [1pt]
 Spatial scale & 0\arcsecpoint66/15$\mu$-pixel & 0\arcsecpoint85/15$\mu$-pixel \\ [1pt]
 Dispersion (at \ha) & 1.019 \AA/15$\mu$-pixel & 1.515 \AA/15$\mu$-pixel \\ [1pt]
 Reduced log dispersion & 48.5 \kms/bin & 75 \kms/bin \\ [1pt]
 Resolution (FWHM) & $\sim$ 2.7 \AA\   & $\sim$ 4.5 \AA\  \\ [1pt]
 Detector & \multicolumn{2}{c}{TI~3-phase~800$\times$800~CCD} \\ [1pt]
 Readout noise & 7e$^-$ & 13e$^-$ \\ [1pt]
 Gain (ADU) & 2.6e$^-$/DN & 1.5e$^-$/DN \\ [1pt] 
 Exposure time & 1800~s & 1500~s  \\  [1pt]
 Net Instrument Eff.$^a$& $\sim 15\%$ & not measured \\ [1pt]
\hline
\multicolumn{3}{l}{\scriptsize a) Net Instrumental Efficiency (telescope, spectrograph, detector) 
 measured at $\sim$ \ha.} \\
\multicolumn{3}{l}{\scriptsize \ \ \ \ Based on fully-open slit observations of 
 spectrophotometric standards.}
\end{tabular}
\end{table}

The nominal exposure times (Table 1) were chosen to sample at least the maxima of the 
rotation curve on both the approaching and receding sides,
but do not allow for measurements at very faint light levels. Typically, the rotation curve
ends just beyond the optical radius (defined here as 3.2 disk scale lengths - see Fig. 4).
Most galaxies were fully sampled within the 2\arcmin slit. 


As in Paper I, the bulk of our reductions was based on the interactive image-reduction 
package VISTA\footnote{VISTA is currently maintained by Jon Holtzman at New Mexico
State University.  The official release is available via anonymous FTP to vista@nmsu.edu.}
originally developed at Lick Observatory (Lauer, Stover, \& Terndrup 1983).  The reduction
procedures described below are available from the author as VISTA routines.  

Red spectra of spiral galaxies at low redshifts typically show the prominent nebular 
lines of \ha$\lambda$6563, \nii$\lambda\lambda$6548,6583, and \sii$\lambda\lambda$6716,6731 
with a relative 
range of intensities.  While nuclear \nii may sometimes exceed \ha in strength
the disk emission is usually stronger at \ha than \nii and \sii.  
The \ha line is also comparable to or more extended spatially than the 
other four emission features.  Addition of these lines would only marginally improve 
the rotation curve solutions, and they were left out for simplicity.  In this
paper, we will thus focus solely on the reduction of \ha RCs. 


\subsection{Basic Procedures and Geometrical Corrections}

Steps were taken to assemble the appropriate 
calibrations to transform CCD frames from a distorted row-column pixel space to an 
orthogonal wavelength-spatial map.  The CCD frames were then re-binned on a regular 
spacing grid and re-sampled such that wavelength and radial increments are constant 
within each bin.
Standard calibration material includes high signal-to-noise dome flats (with same slit
opening as galaxies) and a series of short (bias) and long dark frames. No twilight flats 
were taken. 
He-Ne-Ar comparison lamp exposures were taken at the 
beginning and end of each night, 
and the wavelength calibration zero-point was determined
directly from the night emission skylines.  
The line-curvature was mapped from the calibration spectra whereas ``S-distortion''
mapping was modelled directly from the galaxy's continuum. 

 
\subsection{Preliminary Reductions}


The baseline level (DC offset) was removed as in Paper I.
We also mapped and removed the fixed-pattern noise along the slit direction
due to uniform dark current.
All the images were divided by a map of the pixel-to-pixel sensitivity 
variations in the detector. 
The noise-variance frames described in \S 3.2 were measured prior to flat-fielding to
exclude contributions from the dome, sky, and scattered-light backgrounds. 
This ensures proper error propagation. 
CCD images were also corrected for cosmetic defects.  Hot pixels and bad columns 
were replaced by a local second-order polynomial fit. 
Automatic cosmic-ray rejection requires more delicate attention, especially
near emission lines.  Cosmic rays hits are usually more compact than real spectral 
features but an uneven overlap of two or more such events will easily mimic localized 
emission features.  Cosmic ray removal was thus performs interactively. 

The ``S-distortion'' was finally removed from all the bias-subtracted and flat-fielded 
frames via inverse-shifting of a smooth model from the galaxy continuum.
The amplitude of this distortion at the slit center varied from 1 to 3 pixels 
across the full spectral range.  
The baseline, dark subtraction, flat-fielding, removal of cosmetic 
defects, and correction for S-distortion were all done (in that order)
in batch mode using custom procedures in VISTA.

\subsection{Wavelength Calibration and Line-Curvature}

A polynomial pixel-wavelength fit was extracted from each He-Ne-Ar
calibration spectrum using interactive line identification. Eight to thirteen unblended
lines were identified in each of the lamp spectra, and a third-order polynomial was fit
to derive a dispersion curve.  
The rms residuals in the wavelength solution were of the order of
0.05 pixel, which corresponds to 0.051~\AA~ and 0.076~\AA~(or about 2.5\kms\ at \ha) for
Lick and Las Campanas data respectively.
The precision of most centroiding programs is only about 
1/20$^{th}$ to 1/40$^{th}$ of a pixel, and 0.05 pixel residual is thus satisfactory.  
Barring zero-point flexure shifts, the dispersion solutions were very stable over a run. 
The same calibration lamps were used to model the effects of line-curvature.
The spectra were re-binned along the dispersion axis into a logarithmic wavelength scale 
to provide equal velocity interval per pixel. 
An example of a flat-fielded and rectified spectrum is shown in Fig. 1 for UGC 1437. 



We used position angles listed in the Uppsala Catalog of Galaxies (UGC)
to align the slit of the spectrograph with the galaxy's major axis. These were found 
to be in excellent agreement ($\pm 1^\circ-4^\circ$) with our measured values from 
$r$-band CCD surface photometry (Paper I).

\section{Extraction of RCs}

Extraction of a final rotation curve requires a few more steps;
zero-point adjustment of the dispersion solution for flexure shifts, 
construction of a noise-variance map,
identification of the galaxy's center, sky subtraction, 
and removal of night-sky emission lines and galaxy continuum, 

Tests using a cross-correlation approach 
to determine the center of the galaxy automatically showed that the choice of a 
good template is critical.   The center of bright late-type spirals usually shows
strong lines of \ha and weaker \nii pair.
Since a number of those exhibit \ha/\nii reversal in their center\footnote{The 
strength of this 
effect varies upon the amount of nuclear and disk light which enters the slit aperture 
(Rubin \& Ford 1986, Storchi-Bergmann 1991).  A quantitative model to predict the 
strength of this \ha/\nii reversal is still missing.  The effect has been 
noticed for several decades, but there has not been enough information to choose between 
viable explanations.  Collisional and photo-excitation could play a role, but intrinsic 
variations in core abundances are also possible.  Preliminary results from our data also
show that spirals with bigger bulges and rapidly rising RCs (early-type 
spirals) exhibit stronger ``reversals''.},
we opted for an interactive intensity-weighted marking of the center.
It is still difficult to define a true dynamical center using this technique since
the strength of the emission lines 
does not necessarily correlate with the depth of the central potential.
Moreover, 20\% of the sampled galaxies showed no continuum and/or nuclear 
emission.  In such cases, an eyeball estimate of the center was made. 
A more reliable galaxian center was redefined 
via parametric modeling of the map of velocity centroids (see \S 5).

To quantify and correct for the effects of flexure, the 2D spectra were first 
binned on a linear wavelength scale (with line curvature taken into account) to
extract a spectrum of the night-sky airglow from the galaxy sky background.  The 
zero-point of the dispersion solution was obtained by comparing the skylines of \nei 
and OH transitions with a list of reference wavelengths (Osterbrock \& Martel 1992).
The effects of flexure shifts on the relative measurement of RCs are completely 
negligible but the redshifts themselves are directly affected.  Gravity and 
telescope motion can contribute absolute shifts as great as 4 pixels. 

The skylines and sky background were removed 
by interpolating the regions above and below the galaxy emission with a second-order 
polynomial.  In few cases where the galaxy was too extended and filled the slit, we 
interpolated the sky spectrum from both sides of the \ha emission complex (\ha + \nii). 
An example of skyline and sky background subtraction is illustrated in Fig. 2. 

Last, galaxy continuum or, more generally, all the uniform structure parallel to the 
dispersion axis (\eg stars on major axis) must also be removed.  This operation is 
useful for a closer interactive determination of the galaxian center where continuum
and emission lines intersect.
To do this carefully (especially around the \ha emission complex), we shifted the 
data by the inverse of a coarse (integer) model of the rotation curve so that the 
galaxy emission lines all become straight (to within a fraction of a pixel).  
The background was fitted row-by-row and subtracted from the data. 
The images were then shifted back to recover their exact pre-subtraction 
configuration (see Fig.~5.3).  Integral shifts ensure that no spectrum degradation 
was introduced.  Finally, a noise-variance was constructed for each galaxy spectrum
to allow the measurement of velocity errors (C92). 


We measured the intensity-weighted centroids at each row along the \ha
galaxy emission using a parabolic-binned interpolation (see Appendix C).
This technique makes no assumption about the profile shape and as such, 
yields both a robust center and error estimate. 

The signal-to-noise threshold was chosen to represent an error of a third of a pixel 
in the determination of the centroid.  This corresponds roughly to S/N$<5$ or a 
velocity error greater than 16\kms\ at Lick and 25\kms\ at Las Campanas.  

The final set of CF ORCs is shown in Appendix A.
Figure 4 shows the distribution of the maximal extent of ORCs scaled by
the disk exponential scale length $h$.

\section{Comparison with Other Work}

We now compare our results with the series of photographic ORCs by Rubin and coworkers 
(Rubin \etal 1985, and references therein).
Rubin's sample does not probe the extragalactic volume as deeply as ours and includes 
earlier types.  Consequently, the overlap between our samples is limited to 8 galaxies,
as shown on Fig. 5.  Rubin's smoothed two-sided 
RCs (represented by open squares) were scanned and digitized
from old paper tracings kindly provided to us by Vera Rubin.
Our data (filled circles) are not smoothed and hence display larger fluctuations. 
Most detections (and non-detections) and overall shape variations are well matched.
Our RCs for U1550 and U5250 are slightly shallower
but this is explained by a shorter exposure time than normal.  

Good agreement is also found with 3 galaxies from MAT's collection, as seen in Fig. 6. 

The thesis sample of Vogt (1994) offers a more substantial overlap. 
This is shown in Figs. 7a,b.  Vogt's sample, observed with the Palomar 5m,
was designed to probe the fainter ends of RCs to study environmental 
effects in clusters.  Longer integration times and the aperture ratio both account 
for greater S/N in Vogt's data. 
In most cases though,
the physical extent and broad features between Vogt's data and ours compare
favorably. Typically, detection of nebular \hii\ emission via
long-slit spectroscopy is limited to a common isophotal radius (see Fig. 4)
for nominal TF exposures times ($<2000$s on 5-m class telescopes).
The 23 RCs shown in Fig. 7a,b are reasonably smooth and show little deviation from
circular motions.  Vogt's data show kinematic ``twists'' for 
U1437 and U12631 which are not confirmed by our data (nor are they challenged 
either; here, comparisons suffer from non-detections.)  

Fabry-Perot (FP) spectroscopy offers significant advantages over conventional long-slit 
spectroscopy for the extraction of optical RCs (\cf \eg Schommer \etal 1993).
First, the velocity field is fully mapped spatially. This yields greater S/N in the 
outer regions and provides a more robust measure of the galaxian circular velocity. 
Second, deprojection of the velocities uses kinematical information directly instead
of photometric inclinations for long-slit data which may not be adequate (Schommer 
\etal 1993).  Comparison with the \ha FP data of Amram \etal (1992, 1994) is 
nonetheless quite satisfactory on large scales (Fig. 8).  The FP profiles (open squares) 
consist of a smooth rotation curve projected on the plane of the sky.  
They extend much 
farther out and show less small scale noise than our major-axis cuts (filled circles).
The localized perturbations on the major axis are attributable to spiral arms and
non-circular motions.  Still, we tested that linewidth measurements from smooth fits
of long-slit RCs are in close agreement with similar linewidths from FP ORCs. 
FP data cubes require 
a few hours of on/off integrations on large telescopes ($>4$-m) which makes any FP 
investigation of a large sample of galaxies logistically challenging.  While we recognize
the merits of FP data over long-slit spectra, it is believed that the latter can still
provide sufficient S/N and sampling to measure the most representative value of circular 
velocity for the galaxy.  


\section{Modelling the Rotation Curve}

Before we derive various measures of rotation velocities from ORCs, 
it is worth considering different parameterizations or models which might
best describe their shape.  A disadvantage 
of this approach is that {\it a priori} assumptions about the shape of the RC
must be made.  The major advantage is that all points (and their errors) 
are used simultaneously and systematic deviations in defining a robust 
linewidth measure are likely to be reduced.  
We will see below that ``model'' linewidths prove to be most reliable.

{\it Physical} modelling of RCs is not a simple affair since it requires a complete
description of the contributions from the bulge, disk, and (bright and dark) halo 
components to the overall galaxy dynamics.  Such global models have been proposed, 
first with the Keplerian formulation of Brandt (1960, and references therein) and 
culminating recently in the ``Universal Rotation Curve'' (URC) of disk galaxies by 
PSS which accounts for luminous and dark matter.  One wishes to adopt a 
parameterization which most naturally espouses the shape of each RC, with no pretense
of having a physical basis, yet with the least number of free parameters.  Use of a 
smooth rotation curve model will allow for the measurement of specific rotational 
speeds (metric, isophotal, etc.) and, in principle, extrapolation of the original 
data if they do not extend beyond a certain fiducial radius.  

Three models were tested: 1) a basic 2-parameter $\arctan$ function,
2) a more elaborate {\it ad-hoc} fitting function,
and 3), the ``URC'' parameterization of PSS.  Models 1 and 2 were fitted using
a Levenberg-Marquardt technique.  Application of Model 3 only required input of 
2 variables, the absolute luminosity and the optical radius of the galaxy.
The models were applied to both CF and MAT ORCs (projected on the plane of the sky.)
The fits for all observed CF spectra are reported in Appendix A (on top of the 
actual data), and the fit coefficients for both CF and MAT data are given in 
Appendix B.  Plots and models for MAT's data are available from the 
author upon request.  

It is useful to gauge galaxy sizes in terms of an optical radius and  
we adopt below the definition of PSS for a pure exponential galaxy: 
$r_{opt} \equiv 3.2 h$, where $h$ is the disk exponential scalelength (Paper I). 
More generally, $r_{opt}$ is defined as that radius which encloses 
83\% of the total light.
It is closely equivalent to the isophotal radius at 24 $r$-mag/arcsec$^2$, 
$r^c_{24}$ (Paper I).  We also write $V_{opt} \equiv V(r_{opt})$.

\yyskip

{\noindent {\bf Model 1. \ The ARCTAN function}}

The normalized $\arctan$ RC fitting function is given by:
\begin{equation}
 v(r) =  v_0 + \frac{2}{\pi} v_c \arctan(R),   
\end{equation}
where $R=(r-r_0)/r_t$, $v_0$ is the velocity center of rotation, $r_0$ is the
spatial center of the galaxy, $v_c$ is an asymptotic velocity, and $r_t$ is a 
transition radius between the rising and flat part of the RC. 
This function reproduces the shape of RCs reasonably well with the smallest 
number of arguments.  Moreover, it emerges naturally from the standard 
parameterization of dark halo density profiles
(\cf \eg Gilmore, King, \& van der Kruit 1990, p.212). 
The arctan model is shown in Fig. 9 for various values of
$r_t$ and $v_c = 200$ \kms.  Arctan fits for all galaxies are plotted in 
Appendix A as long-dashed curves on the folded RCs.  The curves are often 
hidden under the continuous line which describes model 2 below.
The arctan model provides an adequate match to most RCs, does not 
account for any sharp peak near the turnover radius (\eg U3181, U7337, U8161).  
It does handle sharp transitions and changing slopes fairly well,
from ``boxcar'' profiles (\eg U10621) to slowly rising curves (\eg U11606, U12304).
The fit parameters for each CF ORC are given in Appendix B, Table A2.1.
A similar list for MAT ORCs is presented in Appendix B, Table A2.2. 
Note that $v_c$ can be arbitrarily large especially for small values of $r_t$. 
Thus, it should not be considered as a useful definition of a galaxy's rotational
linewidth (\cf \eg Schlegel 1995). 

{\noindent {\bf Model 2. \ The Multi-Parameter function}}

A more elaborate, purely empirical formulation is:
\begin{equation}
 v(r) =  v_0 + v_c \frac{(1+x)^{\beta}}{(1+x^\gamma)^{1/\gamma}},   
\end{equation}
where $x=1/R=r_t/(r-r_0)$, and $v_0$, $r_0$, $v_c$, and $r_t$ are defined as
above in Model 1 but $v_c$ and $r_t$ do not take the same values. 
Solid-body rotation, or $v(r) \propto r$ (with $\partial v/\partial r \sim v_c/r_t$),
is recovered for $|r-r_0| \ll r_t$, and flat rotation, or $v(r) \propto v_c$, is
achieved for $|r-r_0| \gg r_t$. 
Similar formulations have also been developed by Schlegel (1995), Rix \etal (1997)
and Kravtsov \etal (1997). 
The term $\gamma$ governs the degree of sharpness of turnover, and $\beta$
can be used to model the drop-off or steady rise of the outer part of the RC, 
mostly for systems with a fast solid-body rise. 
Barring zero-point offsets, this function uses 3 main
fitting parameters: $v_c, r_t$, and $\gamma$.  
The term $\beta$ was used for 30 of the 304 CF galaxies; it is normally 
set to zero in first iterations.  We used only $\beta$ for 16 of 
MAT's 958 modelled RCs.

A representation of typical models is shown in Fig. 10 for a range of 
shape parameters, $\gamma$, at a
given $v_c$ and $r_t$.  The effect of varying $\beta$ is also illustrated
(dotted lines).  For given values of $v_c$ and $r_t/r_{opt}$, 
the multi-parameter fit with $\gamma = 1.1$ reproduces the arctan fit
almost identically.  RCs with larger $\gamma$'s are not as well matched
by the arctan.
The best match of Model 2 to each RC is shown as a solid line in Appendix A. 
The fit parameters for CF and MAT galaxies are again listed in Appendix B. 

%
%

The success of Model 2 is immediately obvious upon visual 
examination of the fits.  Figs. 11a-d show the difference {\chisqr}(mp) - {\chisqr}(at)
between the multi-parameter (Model 2) and arctan (Model 1) fits, against $r_t$ 
and $\gamma$ (both from Model 2) for CF and MAT galaxies.  A preponderance 
of negative differences favors the multi-parameter fitting function; statistical 
equivalence is achieved for $\vert \Delta \chi^2 \vert \leq1$.  
An $F$-test further reinforces the greater confidence level
(of order a few sigmas) of Model 2 over Model 1, especially for MAT galaxies.  
The \chisqr differences are either insignificant or reversed at small values of 
$\gamma$ for rising RCs.  Model 2 also suffers from 
obvious coupling between $r_t$ and $\gamma$,
especially for small values of $r_t$ and $\gamma$.  Fig. 12 highlights
the strong covariance that exists for $r_t/r_{opt} \leq .3$ and 
$\gamma \leq 2.5$.  As expected, slow rising RCs (fainter galaxies)
are favored by 
a model with a low $\gamma$ and high $r_t/r_{opt}$ as seen in Fig. 12.  
Model 1 offers a reasonable match for galaxies with $\gamma \leq 1.5$. 
It is, however, less adequate for the brighter galaxies with greater $\gamma$'s.   
The differences are most striking for rising or declining RCs in their
outer parts and sharply rising curves in their inner parts 
(often associated with strong central concentrations). 

Model 2 fits a greater panoply of RC shapes and shall be adopted below.
Because of internal coupling, Model 2 should be avoided in analyses of galaxian 
structural parameters.
a


{\noindent {\bf Model 3. \ The Universal Rotation Curve (URC)}}

The ``Universal'' parameterization of PSS has recently been used as a tool 
to model TF velocities (G97).  PSS's physical model was designed to account 
broadly for the contribution of the luminous and dark matter to the overall RC.  
In its current form, luminous matter is modelled from the contribution of a stellar 
exponential thin disk and the dark matter halo is expressed as an isothermal sphere.
The dark component is scaled by assuming a maximum (luminous) disk. 
This parameterization implies that all the families of RC shapes can be 
characterized by luminosity alone.  The URC uses two variables as input: $L$, the 
absolute blue luminosity of the galaxy\footnote{We explicitly drop the $B$-band 
subscript for consistency with the PSS notation.}, as the shape parameter and the 
optical radius $r_{opt}$ for normalization at $V_{opt}$.
Equation 14 of PSS is written as:
\begin{equation}
V_{URC}(x) = V(r_{opt}) \left\lbrack \left(0.72  + 0.44\log \frac{L}{L_*}\right)
 \frac{1.97 x^{1.22}}{(x^2 + 0.78^2)^{1.43}} + 1.6 \ e^{-0.4(L/L_*)} \frac{x^2}{x^2
 + 1.5^2(\frac{L}{L_*})^{0.4}} \right\rbrack ^{1/2}  \rm{km \, s}^{-1} 
\end{equation}
with $x = r/r_{opt}$. 

We use $M^*_B = -20.5$, \ie $\log L^*_B = 10.4L_\odot$, in concordance with PSS\footnote{
Both the Stromlo-APM survey (Loveday \etal 1992) and recent results from SSRS2 (Marzke \etal 1997)
for spiral galaxies suggest that $M^*_B = -19.5 + 5 \log (H_\circ/100) = -20.1$, 
or $\log L^*_B = 10.2$ for $H_\circ =75$.
The offset between this and PSS's value will cause model departures at the 2-3\% level.}.

Since derivation of the galaxy luminosity is the final product of the TF machinery,
using a luminosity-dependent formula for TF applications, such as that of PSS,
requires an iterative procedure to solve for the final metric.
We allow ourselves to violate this argument by using measured 
$L_B$ from the RC3 
for the sake of verification.  Color 
transformations to convert available $r$-band luminosities (Paper I) to the $B$-band 
are avoided to minimize errors.  URC models are only computed if both $r$-band and $B$-band 
luminosities are available.
Moreover, these models can only be computed if surface photometry exists to estimate $r_{opt}$,
either directly or from the scale length for pure disk galaxies.  Photometry must also 
extend beyond $r_{opt}$ ($\equiv 3.2 h$).  

The subjectivity in choosing proper scale lengths or applying an exact fraction (83\%)
of the ``total'' light to measure $r_{opt}$ provides an additional source of uncertainty
for those models. Besides being model dependent, $r_{opt}$ increases with wavelength
due to extinction effects (Burstein, Willick, \& Courteau 1994).  Care must thus be 
taken when applying the URC in different wavelength regimes.  We make no 
compensation for dust effects at this point.  

URC models were computed for 131 galaxies and are compared to the data in Appendix A; these
are represented with a dotted curve. 

A visual classification scheme suffices to assess the goodness of fit; 70 models were 
rated ``good'', 21 ``marginal'', and 40 were labelled ``poor''. This degree of 
success ($> 50\%$) is perhaps satisfactory for broad galaxy formation and scaling 
issues but is less adequate for accurate modelling and TF applications (as has been 
recognized by PSS).  The URC fails most conspicuously
for rapidly rising curves and seems to predict too strong a decline in the outer regions. 
In the bulge region, the PSS model has a tendency to overshoot a slow solid-body rise and 
underestimate any fast rise.  This is readily understood by the absence of a bulge term in 
the URC formulation.
While central regions of late-type systems may be small and faint (Courteau, de~Jong, \& 
Broeils 1996), their contribution to the gravitational potential is non-negligible 
(Broeils \& Courteau 1998).  
Furthermore, 
the circular complication involved in using luminosities as input in the URC formula prevents 
any direct TF application.  Issues involving the choice of blue luminosities (which are 
strongly affected by dust), a wavelength- and model-dependent definition (isophotal radii 
or scalelengths),
adjustment of internal constants ($L^*_B$ and $H_\circ$), and revision of the distribution of
light to include a bulge component, all need to be addressed in a more comprehensive analysis.
In its present form, the URC should not be used for exact modelling.  Extrapolation of the URC 
to estimate $V_{opt}$ for shallow RCs is equally admonished in light of the potentially
large data-model differences in the outer parts (\cf Appendix A). 

In what follows, we adopt Model 2 as the best parameterization to describe ORC shapes
for the TF work.  Model 1 may prove more useful for structural analyses. 

\section{Rotational Velocities for TF Use}

The TF relation is based in part on an estimate of the ``intrinsic'' rotational speed, 
or linewidth, of a disk galaxy which is measured beyond the central rise of the RC.
The definition of linewidth varies from author to author.
Linewidth measurements with respect to peak levels of the 21-cm ``horned'' 
profile were initially favored for computational simplicity.
However, peak intensities are affected by sampling and binning effects.
Modern linewidth measurements are based on robust quantities such as 
integrated flux or mean intensity (\eg Bicay and Giovanelli 1986) or the slope of 
the profile edges (\eg Bottinelli \etal 1983). 

Since HI linewidths have defined the standard for most TF calibrations to date,
it is appropriate to seek a new definition of optical rotation which will 
{\it best reproduce 21cm results}.  This new optical linewidth should also meet 
two basic requirements: {\it best internal scatter} 
(\ie robust to observational errors) and {\it minimization of TF residuals.}
A similar search was undertaken by Raychaudhuri \etal (1997) but their small sample
as well as the lack of repeat observations prevented them from discriminating
among their various linewidth measures.  They were able to confirm that
ORCs yield as small a TF scatter as \hi linewidths (see \S 11). 

The measurement of linewidths from resolved RCs is advantageous
since emission line centroids are not affected by internal and instrumental 
broadening.
However, with increasing distance, seeing effects become important and the shape
of RCs can be significantly altered near the center (\cf \eg Vogt \etal 1996).  
At higher
redshifts ($z>1$), the velocity field of a distant galaxy may completely fill the
entrance aperture of either long-slit spectrographs or integral field units. 
It is thus important to keep track of calibrations based on integrated
rotation profiles as well as resolved RCs to allow a bridge between TF calibrations 
at low and high redshifts.

\subsection{Linewidths from Resolved Rotation Curves}

Various measures to estimate the maximum rotational speed from an asymmetric and
noisy rotation curve are investigated below.  Those based on only a few velocity 
points are expected to reproduce rather poorly.  C92
tested 3 linewidth measurements, \dv$_{max1}$, \dv$_{max2}$, and \dv$_{max3}$, 
based on the 1st, 2nd, and 3rd highest velocity points on
each side of the RC.  An average of these, \dv$_{123}$, was also computed
(see \S 7). 

\subsubsection{Weighted Mean Method}

A less susceptible approach consists of taking the weighted mean of all the velocity 
points where the RC levels off.  We do this by summing 20\% of all the velocity 
points from the far-end of the ORC on each side.  For a fast rotator, this 
is equivalent to drawing a line through the points on the rotation plateaus.  The 
mean of the summations on both sides is \dv$_{box}$.  R97 describe this method in
their \S 4.3. 
This technique suffers mostly from an ill-defined interior boundary of integration. 
The weighted-mean method most closely matches the eye-ball technique 
of MAT to fit optical line widths.  
The subjective nature of this method makes duplication and testing 
of MAT's results virtually impossible. 

\subsubsection{Model Linewidths}

We define more objective and stable measures of TF velocity based on the ORC 
parameterization (Model 2) in conjunction with, in some cases, photometric 
parameters.  Four measurements were examined: i) $V_{max}$, 
the peak velocity of the model, ii) $\v22$, the velocity interpolated at 
$r_{disk} = 2.15$ disk scale lengths which corresponds to the location of peak 
rotational amplitude for a pure exponential disk
(\cf \eg Freeman 1970, Binney \& Tremaine 1987, Courteau \& Rix 1997),
iii) $V_{opt}$, the velocity measured at $r_{opt}$, and iv) $V_{iso}$,
an ``isophotal'' velocity inferred at the 23 mag/arcsec$^2$ isophotal level
(\cf Paper I).  


\subsection{Linewidths from Integrated Velocity Profiles}

We now explore alternative linewidths from integrated profile which 
resemble single-dish 21cm spectra.  This is done by adding all the 
velocity points which define the ORC along the spatial axis of the galaxy.  
The motivations for studying this technique are twofold: 1) its principles are
based upon a well-proven method (radio linewidths), and 2) a calibration 
(intensity-weighted) will be necessary for investigations of unresolved
rotation profiles from high-redshift galaxies (Courteau, Sohn, \& Faber 1998). 
The most important benefit of histogram measurements is the increase in linewidth 
S/N since {\it all} the velocity points are used in the computation.

Two velocity histograms are of potential interest: the {\it un-weighted} and 
{\it intensity-weighted} cases.  
The major weakness of {\it intensity-weighted} profiles is that 
a flux integral will favor the brightest \hii regions (Schommer \etal 1993, R97).
This effect is still sufficiently mild in normal spirals (\ie non-starbust, 
LINERs, or Seyfert galaxies) that a TF calibration can still be defined (\eg C92 
and below). {\it Un-weighted} velocity histograms alleviate this shortcoming entirely. 

\subsubsection{Un-Weighted Histograms and Probable Linewidths}

One can sort all the velocity data points (which obey the minimum S/N criterion) 
and calculate the difference, $W_{hist}$, between the $10^{th}$ and $90^{th}$
(interpolated)
percentile points in the ranked distribution of velocities.  This technique,
first used by Dressler \& Faber (1990) and further developed by Vogt (1994), is 
similar to constructing an unweighted velocity histogram and measuring the 
width at 10\% of total area on each side of the profile.  
A variant of this technique called the ``Probable Min-Max'' (PMM) method 
by R97 (see their \S 4.6), considers the probability distribution of all 
velocity points with their Gaussian errors. 
A velocity $W_{prob}$ is computed from the difference of the 
90\% probable maximum and minimum from both sides of the RC. 
This method is however sensitive to broadening effects which are reflected in 
the velocity error at each point. 

\subsubsection{Intensity-Weighted Histograms}
We explored three flux-weighted linewidth measurements: $W_{15}$, $W_{20}$, 
and $W_{25}$.  
The area under each profile is measured by integrating along the 
velocity axis; fractional flux levels are computed via parabolic-binned 
interpolation (see Appendix C). 
We define a velocity linewidth, $W$, as the difference between those
velocities where the integration reaches a predefined percentage of 
the total area, \eg 10\% and 90\% for $W_{20}$.
Definitions relative to peak intensities are avoided since the detailed shape 
of the profile is sensitive to spectral resolution and, for ORCs, to the individual 
fluxes of \hii\ regions lying on the slit.  While the exact shape of the profile 
is not a conserved quantity, its area or total flux, should be.
Fig. 13 shows the locations of these width markers for UGC1437. 
These are measured quite high compared to 21cm profiles (\cf \eg Roberts 1978, Bicay \&
Giovanelli 1986).  A discussion of this choice of optical linewidths and systematic 
effects will be given in Courteau, Sohn, \& Faber (1998).  

We consider one last measure of integrated linewidth, $W_{50m}$, the separation
between the velocity points where the intensity reaches 50\% of the mean flux over the full 
velocity range of detectable signal.  This is analogous to 21cm linewidths
defined at 50\% of the total mean. 

Rotation profiles for each galaxy were not presented for space considerations but
are available at request from the author. 



\subsection{Logslopes}

For future use, we also compute the logarithmic slope, $d\log v(r)/ d \log r$, 
from the folded deprojected data in the rest-frame of the galaxy. 
Measurement of logslopes from sparse data is exceedingly sensitive to noise
and we use a bootstrapping technique\footnote{We use linear least-square 
fitting on a random selection of all the velocity points included in the 
fit.  This operation is 
repeated 100 times.  A loglsope is computed if at least 6 velocity points are 
available.  Otherwise, it is set to zero.} to infer their mean and error. 
The mean logslope of the observed data, $ls_r$, is computed between $r_{disk}$ 
and the last measured velocity point.  We also compute an analogous quantity,
$ls_m$, from Model 2.  These are given in Table 13. 

Fig. 14 displays the distribution of logslopes, $ls_r$ and $ls_m$, for all CF 
galaxies.  About 4\% of the sample exhibits declining outer slopes and $\sim$ 60\%
of all RCs have achieved convergence, \ie $ls_r \leq 0.2$. This fraction of 
convergent profiles is lower than the 81\% found by Schlegel (1995) for a sample
of 282 nearby IRAS-selected galaxies.  None of Schlegel's RCs exhibit negative 
logslopes either. 
This high rate of convergence of IRAS-selected galaxies, if true, is surprising
since the TF scatter for these galaxies was found to be higher than for optically-selected
spirals (Schlegel 1995).  One would expect that better behaved linewidths (\ie more convergent 
profiles) would yield 
reduced TF scatter.  Massive star formation activity characteristic of IRAS galaxies could 
however bias the luminosity high and explain Schlegel's observations. 
van~Driel \etal (1995) also studied TF differences between optically-selected 
and IRAS galaxies but did not find striking discrepancies. 
Further investigations of 
IRAS-galaxies would be desirable, especially if their RCs are systematically 
flatter than ORCs as indicated by their logslopes. 
 
\yskip
\section{Internal Errors from Repeat Measurements}
\label{int_errors}

We have measured repeatability errors from multiple observations of 62 galaxies,
for a total of 132 independent observations.  These are all reported in Appendix A.
Unweighted standard deviations, $\left<\sigma^2_{obs}\right>^{\frac{1}{2}}$, in \kms\ 
are given in Table 2 below for the various linewidth measures. 
These refer to the one-sided value of the rotational velocities (\ie \dv$/2.$)

Not surprisingly, all measures of linewidth based on one or a few velocity points fare 
rather poorly.  
The ``box'' method shows improved internal errors purely for statistical reasons. 
This technique is equivalent to
fitting the best horizontal line through the flat rotation curve.

Internal errors for all model linewidths and the velocity histograms
are small and hold great promise for TF analysis.
Errors above 8 \kms\ are unsatisfactory and suggest that the
linewidth measure is too sensitive to observational errors. 
Of the linewidths with smallest internal scatter, $\v22$, $V_{opt}$, and $V_{iso}$, 
are however photometry-dependent.  
The linewidth with best internal scatter, $W_{hist}$, also holds the distinction of
being completely independent of any modelling of the RC or scaling via 
photometric parameters. 
Its operational simplicity also makes it ideal for widespread use.

\yyskip
\begin{table}[t]
\centering
\tablenum{2}
\begin{tabular}{rccccccc}
\multicolumn{8}{c}{Table 2} \\
\multicolumn{8}{c}{RMS Error of a Single Observation for {\it Raw} Velocity Parameters} \cr
\hline\hline
 & \dv$_{max1}$ & \dv$_{max2}$ & \dv$_{max3}$ & \dv$_{123}$ & \dv$_{box}$ & $V_{max}$ & $\v22$ \cr
 std. dev.:  & 15.3 & 8.5 & 7.7 & 10.0 & 7.0 & 6.6 & 5.7 \cr \hline
 & $V_{opt}$ & $V_{iso}$ & $W_{hist}$ & $W_{prob}$ & $W_{15}$ & $W_{20}$ & $W_{50m}$ \cr
 std. dev.:  & 5.7 & 5.6 & 5.5 & 8.3 & 5.9 & 6.3 & 12.6 \cr \hline\hline
\end{tabular}
\end{table}
\yyskip

The three raw linewidths, $W_{15}$, $W_{20}$, and $W_{25}$ all show 
internal scatter comparable to those quoted for radio velocity widths (Bothun
\etal 1985, Freudling 1990, Mould~\etal 1993).  
The errors on $W_{50m}$, which samples the optical profile lower in the wings, 
are too large to make it a good velocity width indicator.  
We found no correlation between linewidth error and distance (redshift) or
absolute luminosity (linewidths). 

\section{Corrections to Linewidths}

Observed line-of-sight velocity widths, whether they are measured from 
resolved RCs or integrated profiles, must be corrected for projection 
on the sky and cosmological stretch. 
The corrected rotation width, $W_c$, is 

\begin{equation}
  W_c = \frac{W_{raw}}{(1+z) \sin i}
\end{equation}

{\noindent where the observed {\it raw} linewidth, $W_{raw}$, must be treated for 
various broadening effects prior to deprojection {\it if} the widths are 
derived from an integrated spectrum. 
A potential source of error for resolved linewidths is that of slit misalignment. 
To test this, we rotated the slit at various position angles spanning a range
of $8^\circ$ on a few test galaxies.  Results showed no discernible effect.
Giovanelli \etal (1997) also simulated this effect and report negligible 
errors for PA offsets less than $15^\circ$.}


Corrections to integrated profiles must account for broadening effects such
as instrumental resolution and focus variations, ``turbulent'' (non-circular,
perpendicular) motions of the gaseous component, signal-to-noise of the spectrum, 
and smoothing. 
Instrumental resolution
and focus variations were accounted for by measuring
the width of OH emission-lines 
at the same height as that measured on the galaxian integrated profiles
(see Table 1 for representative values.)


Since the full width is a convolution of the true profile
of the galaxy with that of the instrument, we assume that

\begin{equation}
\rm{W_{raw}^2} = \rm{W_{galaxy}^2} + \rm{W_{resolution}^2}
\end{equation}

This is strictly true for Gaussian profiles, but tests show that it seems to work 
adequately for double-peaked profiles as well. 
The typical contribution of instrument and focus broadening to the final width
is $\sim 5\%$ for fast rotators ($V_{max} > 200$ \kms) and $\sim 9\%$ for slow rotators
($V_{max} \leq 200$ \kms).  Focus variations ($\sigma_f \sim 10$ \kms) 
contribute less than a percent to the overall width.

A correction for turbulence was derived by Tully \& Fouqu\'e (1986; 
hereafter TF86), based on the earlier treatment of Bottinelli \etal (1984).  Eq. 12 from
TF86 assumes that the raw linewidth, $W_{20}$, is already corrected for instrumental 
resolution effects.  
Application of this formula would not be strictly adequate here since we 
have not demonstrated 
equivalence between their $W_{20}$ and our linewidths corrected for instrumental broadening. 
Independent investigations (Rhee \& Broeils 1997) suggests that TF86's formulation 
slightly overestimates the broadening effects of turbulence (which itself is a rather 
poorly constrained phenomenon), but regardless of the exact treatment, this correction 
is small for the large galaxies considered here.  We chose to neglect it in our treatment.
Moreover, unlike instrumental broadening, turbulence does not affect the TF scatter
and therefore, should not alter our conclusions. 


The inclinations were estimated using CCD ellipticities, $\varepsilon$, 
and assuming an intrinsic flattening ratio, $q_\circ$,
of 0.18 for all the galaxies (see Paper I for a discussion on the choice of ellipticities).
We use
\begin{eqnarray}
   \cos^2 i(\varepsilon) = \frac{(1-\varepsilon)^2 - q_\circ^2}{1 - q_\circ^2}.
\end{eqnarray}
where the ellipticities for each galaxy are given in Table 4.
Slightly different estimates for the flattening parameter, $q_0$, are 
quoted in the literature; for disk galaxies, these vary between 0.11 to 0.20 (see 
\eg Haynes \& Giovanelli 1984, B85, Han 1991, Willick 1991, Rhee 1996, G97).  Variations 
in the choice of $q_\circ$ have very little effect on the projection correction of the 
velocity width (col. 3) due to the slow variation of the sine function for large 
inclinations (where ($1 - \varepsilon$) becomes
comparable to $q_\circ$); the inclination-dependent correction for internal extinction
is also very little affected by the precise choice of $q_\circ$ since it is assigned a fixed
value for $i >$ 80\sdeg\ (see Paper I). 
For $\varepsilon \geq 0.82$, where inclination
measurements become limited by the intrinsic thickness of the disk, the inclination value
was set to 90\sdeg.
Inclinations are in principle affected by atmospheric seeing (\eg Saglia \etal 1993, G97)
but the seeing disk is small compared to the size of our galaxies.  Our sample is limited 
to $i \leq 75\sdeg$ where seeing effects are reduced. 


 
\section{Presentation of the Data}

We present in Table 3 a list of the main measured spectroscopic quantities for all the CF
galaxies. In cases where a galaxy was observed more than once, the measurements are an 
unweighted average. 
All the velocity widths
are corrected for inclination and redshift broadening following Eq. 4. 
Integrated linewidths (only $W_{15}$ and $W_{20}$ are presented here) are also corrected
for instrumental and turbulent broadening.  All the linewidths show twice the galaxian 
rotation.  In cases where {\it model} linewidths could not be measured due to shallowness
of the RC, the velocity is listed as zero. 
Recall finally that logslopes are computed from deprojected velocities. 
Only the first page of this table is shown here. 
Table 3 is arranged as follows:
\begin{description}
\item[{\sl Col. (1):}] UGC number of the galaxy. The last two entries refer to CGCG
    reference numbers (\cf Paper I for PCG numbers);
\item[{\sl Col. (2):}] $n$, the total number of observations per galaxy;
\item[{\sl Col. (3):}] $i$, the inclination of the galaxy in degrees calculated
     with Eq. 6 from ellipticities given in Paper I.  An intrinsic flattening
     of 0.18 is assumed; 
\item[{\sl Col. (4):}] $cz_{\rm LG}$, the radial velocity corrected for motion 
 with respect to the Local Group, $cz_{LG} = cz_\odot + 300 \sin(\ell) \cos$(b).
 The heliocentric redshift, $cz_\odot$, is defined as the center of symmetry of the 
 rotation curve as measured with Model 2 (\cf Paper I); 
\item[{\sl Col. (5):}] $cz_{\rm CMB}$, the radial velocity in the frame of the Cosmic
     Microwave Background.  The correction uses the motion of the Sun with respect to
     the CMB frame determined by the COBE dipole anisotropy (Kogut \etal 1993);
\item[{\sl Col. (6):}] $W_{15}$, the velocity width measured from the flux-weighted 
     rotation profile at 15\% of the total area;
\item[{\sl Col. (7):}] $W_{20}$, same as $W_{15}$ but measured at 20\% of the total area;
\item[{\sl Col. (8):}] $V_{max}$, the peak velocity from the fitted RC with Model 2;
\item[{\sl Col. (9):}] $\v22$, the velocity from Model 2 measured at $r_{disk} = 2.15h$, where 
    $h$ is the disk scale length from 1D bulge-to-disk decompositions (\cf Paper I);
\item[{\sl Col. (10):}] $V_{opt}$, the velocity from Model 2 measured at $r_{opt} = 3.2h$, the optical radius
    defined as in PSS; 
\item[{\sl Col. (11):}] $W_{hist}$, the velocity difference between the $10^{th}$ and $90^{th}$
    (interpolated) percentile points in the ranked distribution of velocities;
\item[{\sl Col. (12):}] $W_{prob}$, the velocity difference between the probable minimum and
 maximum of the RC at the 90\% level (\cf R97); 
\item[{\sl Col. (13):}] $ls_m$, the logslope measured from Model 2
    between $r_{disk}$ and the last fitted velocity point; 
\item[{\sl Col. (14):}] $ls_r$, the logslope measured from the observed RC 
    between $r_{disk}$ and the last measured velocity point; 
\item[{\sl Col. (15):}] $D$, the distance of the galaxy in Mpc, 
     $D = cz_{\rm LG}$/\hub; 
\item[{\sl Col. (16):}] $r_{max}$, the maximum spatial extent in arcseconds of 
     either the approaching or receding side of the galaxy spectrum; 
\item[{\sl Col. (17):}] $r_{kpc}$, same as above but in units of kpc. 
\item[{\sl Col. (18):}] $m_c^r$, the corrected $r$-band total apparent magnitude
     from Paper I, which here also includes a small $K$-correction term (\cf
     Willick \etal 1997, Eq. A14); 
\item[{\sl Col. (19):}] $M_c^r$, the corrected absolute $r$-band magnitude.
     $M_c^r = m_c^r - 5 \log D[{\rm Mpc}] + 25$; 
\item[{\sl Col. (20):}] Ft, the Freeman type (1 or 2) of the galaxy (Freeman 1970).
\end{description}

\notetoeditor{Table3 (Courteau.tab3.tex) in LANDSCAPE format should
 be inserted here}

\section{Optical TF Relation}

Below, we apply ``forward'' linear regressions to the TF equation $M_c^r = a\eta_c + b$
 for 8 measures of optical linewidths, where $\eta \equiv \log V_c - 2.5$.
We are now concerned specifically with finding the linewidth definition which yields the
smallest TF scatter.  The correct derivation of the forward TF slope and zero-point 
requires that magnitudes be corrected for selection bias (W94, W96).  Here we limit 
ourselves to relative differences where the exact bias treatment is not important.

The complete CF sample includes regions of density enhancements near the Perseus-Pisces
(PP) Supercluster that must be excluded to avoid any further bias of the TF scatter.
In C92, we identified a region of Quiet Hubble Flow (QHF) away from the influence of the 
Local Supercluster which is nearly orthogonal to the main direction of flow between PP
and the Great Attractor (\cf also Courteau \etal 1993).  The Coma Supercluster and the 
Great Wall
are included in the QHF but no large-scale velocity features are visible there.  Indeed, the 
mean infall signal in the vicinity of the Great Wall and the measured shear across it 
are both consistent with zero (C92, Bernstein \etal 1994, Dell'Antonio, Geller, \& Bothun 1996).
A measurement of TF scatter in the QHF region will thus provide a lower limit to the real
TF scatter.  This is especially true since bias corrections are neglected here.  
Such corrections tend to amplify scatter estimates (W96).  A similar QHF control region can 
be defined for MAT's sample (Mathewson \& Ford 1994).  However, the infall pattern due to
the Great Attractor is not as clearly delineated there as at PP and differences between 
TF relations in the Southern QHF and GA regions are hardly noticeable.

We use linear least-square fitting
to compute values of the TF dispersion for the full sample 
of 304 CF galaxies and its subset of 127 QHF galaxies. 
Values for $W_{15}$, $W_{20}$, $W_{25}$, $W_{hist}$, $W_{prob}$, and $V_{max}$ exist 
for all galaxies but $\v22$, $V_{iso}$, and $V_{opt}$ may be missing for small
systems (\cf \S 6.1.2). 
Fig. 15 shows TF relations for 6 combinations of fully corrected magnitude and 
linewidths.  The open circles show all CF galaxies; those filled are specific to 
the QHF region only. 
A similar exercise was done for the 958 MAT galaxies and a subset of 310 QHF galaxies.
Fit coefficients for both samples are reported in Table 4 below.  The fits for MAT 
are as above
except that $W_{15}$ and $W_{20}$ were replaced by $V_{rot}$, the advocated linewidth
by MAT for final TF analysis.  $V_{iso}$ linewidths were not measured for MAT data.
The TF plots for that sample (not shown here) are similar to those of CF (\cf W97).


\begin{table}[th]
\centering
\tablenum{4}
\begin{tabular}{l | c c c c c c c}
\multicolumn{8}{c}{Table 4} \\ 
\multicolumn{8}{c}{TF fits} \\ \hline\hline
& \multicolumn{3}{c}{ALL CF galaxies} & & \multicolumn{3}{c}{CF QHF galaxies} \\ \cline{2-4} \cline{6-8}
& $a\,(\pm)$ &\multicolumn{1}{c}{$b\,(\pm)$}& $\sigma_{TF}$(mag) & & 
  $a\,(\pm)$ & $b\,(\pm)$ & $\sigma_{TF}$(mag) \\ \hline
$W_{15}$   & -6.34(0.24) &  -20.74(0.03) & 0.52 & & -5.89(0.32) & -20.74(0.04) &0.41  \\ [1pt]
$W_{20}$   & -6.35(0.24) &  -20.74(0.03) & 0.51 & & -5.96(0.39) & -20.75(0.04) &0.39  \\ [1pt]
$W_{hist}$ & -5.77(0.22) &  -20.72(0.03) & 0.51 & & -5.40(0.28) & -20.74(0.04) &0.39  \\ [1pt]
$W_{prob}$ & -6.66(0.26) &  -20.26(0.04) & 0.53 & & -6.03(0.32) & -20.35(0.05) &0.40  \\ [1pt]
$V_{max}$  & -6.09(0.26) &  -20.62(0.03) & 0.55 & & -5.34(0.36) & -20.69(0.05) &0.46  \\ [1pt]
$\v22$ & -6.36(0.22) &  -20.77(0.03) & 0.46 & & -6.17(0.28) & -20.77(0.03) &0.34  \\ [1pt]
$V_{iso}$  & -6.47(0.25) &  -20.70(0.03) & 0.45 & & -6.02(0.31) & -20.69(0.04) &0.36  \\ [1pt]
$V_{opt}$  & -6.99(0.33) &  -20.60(0.04) & 0.46 & & -6.92(0.42) & -20.59(0.05) &0.34  \\ \hline\hline
& \multicolumn{3}{c}{ALL MAT galaxies} & & \multicolumn{3}{c}{MAT QHF galaxies} \\ \cline{2-4} \cline{6-8}
$V_{rot}$   & -6.91(0.11) &  -21.37(0.02) & 0.61 & & -6.52(0.15) & -21.43(0.03) &0.58  \\ [1pt]
$W_{hist}$  & -6.51(0.10) &  -21.46(0.02) & 0.57 & & -6.62(0.17) & -21.49(0.03) &0.55  \\ [1pt]
$W_{prob}$  & -6.83(0.12) &  -21.13(0.02) & 0.59 & & -6.95(0.20) & -21.15(0.03) &0.57  \\ [1pt]
$V_{max}$   & -6.68(0.13) &  -21.34(0.02) & 0.64 & & -6.92(0.20) & -21.36(0.03) &0.62  \\ [1pt]
$\v22$      & -6.64(0.13) &  -21.50(0.02) & 0.56 & & -7.14(0.19) & -21.52(0.03) &0.51  \\ [1pt]
$V_{opt}$   & -6.76(0.22) &  -21.41(0.03) & 0.59 & & -6.38(0.34) & -21.46(0.05) &0.53  \\
\hline
\end{tabular}
\end{table}

Slopes are of order 6.0 for the CF sample which is somewhat shallower 
than those reported in W97, regardless of bias correction. 
This stems from slightly improved linewidths and magnitudes. 
The MAT data show larger TF scatter than reported in W97; this is due to our 
use of MAT's original magnitudes for heuristic reasons (improved magnitudes 
were derived in W97).

Local infall motions clearly inflate the TF scatter in
the full CF sample.  Both CF and MAT QHF samples show a tighter dispersion at all 
linewidths; this is especially true for the CF sample. 
As seen in Fig. 15, the scatter in magnitudes is essentially constant at all 
absolute luminosities for resolved {\it model} velocities yet it does increase
slightly at the fainter (smaller) end for {\it histogram} linewidths.  
The possibility of curvature in this sample has been discussed in W97
but effects are weak.  Histogram linewidths may exhibit more of a curved 
trend than model linewidths but statistics are too poor to rule in favor
of a quadratic transformation. 

``Isophotal'' linewidths, $\v22$, $V_{iso}$, and $V_{opt}$,
yield the smallest TF residuals.  These also showed lowest 
internal scatter for the CF sample (\S 7), though repeatability does 
not necessarily ensure scatter tightness in these TF correlations. 
For example, $W_{hist}$ and $W_{20}$ repeat with quite different accuracy yet
they yield rather similar TF solutions, with identical $\sigma_{TF}$'s
for the CF sample.
$V_{max}$ also fares very poorly.  Its internal error is 
satisfactorily low, yet as a TF observable, it yields the worst TF solution,
for both samples.  The most favorable definition of TF linewidth so far, 
as inferred from both the CF and MAT data sets, is that of $\v22$.  

Similar tests were performed by R97 with 25 galaxies in the vicinity of Coma. 
Their ``Weighted Mean Method: Outer Segments'' is equivalent to our $V_{box}$,
and both techniques yield comparably poor results (not shown in Table 4). 
Their PMM measure is also identical (by definition) to our $W_{prob}$, yet their
result seems more optimistic than ours.  Our investigation yields 
a moderate-to-poor TF scatter for $W_{prob}$ while R97 rate it as their 
second best.  R97's best measure actually comes from a weighted mean of the velocity
points in the range $\sim 0.5-2 r_{opt}$.  This is a surprising measure of rotation 
since it extends beyond the normal reach of most ORCs (\cf fig. 4).
Thus we are unable to reproduce their result. 
R97 also showed that velocities measured at $1.3 r_{1/2}$ yields a poor RMS TF scatter.
For an exponential disk, this is equal to $\v22$!
Our results appear to differ from R97's in most respects (though error bars in R97's 
$\sigma_{TF}$s' are larger since their sample is ten times smaller than CF).  
R97 intentionally avoided model fitting and therefore comparison with the majority 
of our model linewidths cannot be made.

It should also be mentioned that our lowest value of TF scatter (0.34 mag) compares 
favorably with mean dispersions from large TF calibrations based on \hi linewidths (G97). 

\section{Correlation with \hi}

As a final test for linewidth selection, we examine which of 
the tested measures of rotation best correlates with 21cm linewidths.  
A correlation between optical and radio velocity widths is expected 
on a dynamical basis.  However, even well-developed ORCs rarely extend beyond 
$1.5 r_{opt}$ whereas \hi RCs often reach as far as 3 to 5 $r_{opt}$ (Broeils 1992)
(see Fig. 16). 

Various calibrations of optical/radio linewidths have already been established. 
Thonnard (1983) found that 21cm linewidths measured at 50\% of peak intensity
yield a tight correlation: $\left< V_{opt}^{max} - W_{21cm}^{50} \right>
= -1.5 \pm 3.9$ \kms. 
C92 compared line-of-sight rotational velocities for 205 CF galaxies
with radio linewidths from the compilation of Giovanelli and Haynes (private 
communication 1990).
A tight relationship was established with a slope of 1, minimal zero-point,
and $\sigma_{opt/21cm} = 24$ \kms\ 
(\cf C92, Fig. 5.9).  In this correlation, 
the scatter was found to be remarkably constant over the full range of
measured velocities ($100 \leq V_{max} \leq 350$ \kms).
MAT showed the same trend and tight scatter from 75 to 290 \kms\ 
for 219 galaxies (\cf their Fig. 5).
Their best measured dispersion is $\sigma_{opt/21cm} = 20$ \kms.  A similar calibration 
by Vogt (1994) for 96 galaxies yields $V_{opt} = 1.02 W_{21cm} - 4$ \kms\ with 
$\sigma_{opt-21cm} = 30$ \kms.  From Table 1 of R97, we also compute 
$V_{opt}^{\rm PMM} = 0.95 W_{21cm}^{50} - 5.76$ \kms\ ($1\sigma=13.41$ \kms) and
$\Delta V_{opt}^{0.5-1r_{opt}} = 0.95 W_{21cm}^{50} - 29.71$ \kms ($1\sigma=16.8$\kms).
Although the canonical linewidth scatter is $\sim$ 25-30 \kms, the smaller values quoted 
by R97 stem mostly from the quixotical degree of pruning of their sample.
Linewidths listed in their Table 1 do not include deprojection
corrections, which also yield lower dispersion estimates. 

To test our optical velocities against 21cm linewidths, we restrict our comparison
to the recent list of Giovanelli, Haynes \& coworkers (G97). 
This collection is the largest and most uniform compilation of published 
21cm linewidths to date.  Though G97 do include linewidths from various sources, we
use only those directly measured by them.  These widths 
are measured at 50\% intensity of the profile horns.

Our comparisons use {\it fully corrected}\ linewidths to assess the maximum 
difference due to various transformations and sources of errors between final 
TF variables.  
We have derived the following linear and logarithmic 
transformations: $V_c = A W_{GH} + B$ and $\log V_c = A^\prime 
\log W_{GH} + B^\prime.$  Transformation coefficients are given in Table 5 below.


\begin{table}[th]
\centering
\tablenum{5}
\begin{tabular}{l | c c c c c c c}
\multicolumn{8}{c}{Table 5} \\
\multicolumn{8}{c}{Optical/Radio Linewidth Transformations} \\ \hline\hline
& \multicolumn{3}{c}{Linear Fit} & & \multicolumn{3}{c}{Logarithmic Fit} \\ \cline{2-4} \cline{6-8}
& $A\,(\pm)$ &\multicolumn{1}{c}{$B\,(\pm)$}& $\sigma$ & & 
  $A'\,(\pm)$ & $B'\,(\pm)$ & $\sigma'$ \\ \hline
$W_{15}$   & 0.98(0.06) & \phantom{0}-9.5(22.6) & 41.6 & & 0.969(0.080) & \phantom{-}0.053(0.202) & 0.063 \\ [1pt]
$W_{20}$   & 0.98(0.06) &           -10.0(22.1) & 40.7 & & 0.974(0.077) & \phantom{-}0.040(0.195) & 0.060 \\ [1pt]
$W_{hist}$ & 1.10(0.07) &           -42.7(24.6) & 45.2 & & 1.199(0.090) &          -0.526(0.228) & 0.070 \\ [1pt]
$W_{prob}$ & 1.13(0.07) & \phantom{0-}0.8(25.8) & 47.3 & & 1.210(0.135) &           0.068(0.197) & 0.053 \\ [1pt]
$V_{max}$  & 1.01(0.09) & \phantom{0}-0.5(31.1) & 57.2 & & 0.990(0.128) & \phantom{-}0.019(0.322) & 0.099 \\ [1pt]
$\v22$ & 0.99(0.05) & \phantom{0}-0.1(17.0) & 29.2 & & 0.988(0.062) & \phantom{-}0.019(0.155) & 0.044 \\ [1pt]
$V_{iso}$  & 1.00(0.06) & \phantom{0-}4.8(20.7) & 33.6 & & 1.003(0.074) &          -0.006(0.183) & 0.048 \\ [1pt] 
$V_{opt}$  & 0.99(0.08) & \phantom{0}20.3(29.2) & 39.1 & & 0.892(0.107) &           0.294(0.273) & 0.047 \\ 
\hline
\end{tabular}
\caption[Fit coefficients for linewidth transformations]{\label{tab:Dvfits}
  Fit coefficients for linewidth transformations from the system of G97. 
  The standard
  deviations and zero-points of the fits are expressed in units of \kms\ or log({\kms}).}
\end{table}

The optical rotational velocity which matches 21cm linewidths best, both in linear and
log space, is again $\v22$.  Its linear transformation slope is essentially one
and the zero-point is consistent with zero, indicating a virtual one-to-one 
correspondence between $\v22$ and the radio linewidth $W_{21cm}^{50}$ of G97. 
$V_{max}$ is again the worst contender.  Our
linear transformation for $W_{hist}$ is also
much poorer than that of Vogt (1994) who uses a similar algorithm.  The zero-point
offset is widely different and we find a larger dispersion (by 10\kms). 
The data from R97 also favor the PMM linewidth over a weighted-mean linewidth 
($\Delta V(0.5-1r_{opt})$) as a better match to radio linewidths, yet the latter
yields a tighter TF scatter.  No explanations for the trends seen by Vogt and R97 
are offered at present.
We contend for now that our {\it model} linewidths $\v22$ and $V_{iso}$ provide 
the closest match to the \hi linewidths of G97. 

The tight and constant scatter at all luminosities between (suitably chosen) 
optical/radio linewidths (C92, MAT, Vogt 1994) may appear surprising given that resolved 
\hi RCs extend far beyond the optical radius.  Fig. 16 shows series of re-scaled
\hi RCs from Broeils (1992) to illustrate this point.  The arrow indicates 
the location of the optical radius.  As seen in panels b) and c) (most appropriate for
comparison with typical TF galaxies), many \hi RCs keep rising 
beyond $r_{opt}$.  Perusal of Appendix A (see also MAT, Rubin \etal 1989) also 
shows many ORCs which have not reached any turnover by the last 
point of detection (see also Fig. 14).  It is often 
assumed that optical linewidths should be biased {\it low} in comparison to 
radio data, especially in low mass systems. 
%
However, unlike resolved \hi RCs, \hi linewidths {\it do not} sample 
outer disks effectively since the \hi  surface density drops rapidly
beyond the optical radius (Cayatte \etal 1994, Broeils \& van Woerden 1994,
Broeils \& Rhee 1997).  The amount of \hi inside $r_{opt}$ is typically 
greater than $50-60$\% of the total.  This holds true for the majority of spiral
types (Cayatte \etal 1994, Roberts \& Haynes 1994).  
\hi linewidths are thus weighted heavily by the \hi flux inside the optical 
radius.  Hence, it is not surprising that optical resolved velocity widths 
and \hi integrated linewidths should match so well.  
For low and intermediate mass galaxies (Fig. 16a,b), the \hi  profile width 
may therefore be less than twice the true maximum velocity.

Modelling the interplay between the shape of the rotation curve, 
the \hi  surface density, and the width of the global \hi  profile 
is an important task.  The correlation between 21cm linewidths and 
the true maximum rotational velocity of spiral galaxies, often assumed 
to be unity but surely smaller, would be of great benefit for dynamical studies. 
Exploratory investigations are being conducted (\eg Schulman \etal 1997) 
but a clear picture is still lacking.

A significant fraction (30\%) of disk galaxies also reach their peak
velocity close to the optical radius
which reinforces the tight
optical/radio linewidth correlation.  Fig. 17 shows the distributions
of peak rotational velocities from the (resolved) optical and \hi RCs
of CF and Broeils (1992) respectively.  The open circles
represent only those galaxies in CF that have achieved convergence 
(\S 6.3) and Broeils' galaxies are shown as filled circles labelled 
with their catalog name.  
The solid line represents a transition between HSB and dwarf
galaxies.  Above that line, most of the galaxies show a turnover
near $4-5h$ which is just within the limits of optical detection
by conventional long-slit spectrographs. 

\section{Summary}

We have presented a new collection of optical RCs for
304 late-type spirals, here referred to as the Courteau-Faber (CF)
sample.  The first sections of this paper discussed 
data reduction procedures and extraction of RCs from 
resolved \ha\ spectra.  Our rotation curves were shown to be in 
excellent agreement with similar data taken by different authors 
with different telescopes (Amram, Mathewson, Rubin, and Vogt). 
An empirical function was adopted to model the shape, amplitude, 
and zero-point (radial velocity) of each ORC, and  
then used to infer stable linewidths for TF applications.
An additional fitting parameter was sometimes included to account for 
rising or declining RCs beyond the optical radius.  Strong covariances 
are unavoidable in this multi-parameter modelling.  A two-parameter
prescription, like the arctan function, is required to minimize
internal coupling between the fitted parameters.  We also modelled 958 RCs 
from the collection of Mathewson \etal (1992) using the same fitting functions. 

We pointed out that the ``Universal Rotation Curve'' of Persic, Salucci, 
\& Stel (1996) is not a perfect tracer of fast rising bulges and declining 
outer disks.  This shortfall of the model may soon be remedied by accounting
at least for a bulge component (Salucci 1997, private communication). 
Because of matching uncertainties at large radii, use of the URC model
is not recommended to extrapolate isophotal TF linewidths.  The 
model-dependency of the URC via $r_{opt}$ and metric requirement to define 
$L_B$ are additional sources of ambivalence.

The latter half of this paper focused on determining the best measure 
of rotational velocity for TF work.   Our selection was based on three 
criteria:

$\bullet$ Smallest internal scatter (\ie robust to observational errors),

$\bullet$ Minimal TF residuals, and

$\bullet$ Best match to radio (21cm) linewidths. 

The best linewidths are those determined via multi-parameter fitting. 
Modelling smooths out all effects of non-circular motions which are 
particularly acute in long-slit data.  Internal 
repeatability is thus maximized.  The best measure of TF velocity
is $\v22$, the location of peak rotational velocity for a pure 
exponential disk. 
$\v22$ is interpolated from the {\it model} rotation curve at 
$r_{disk} (= 2.15h$, where $h$ is the disk exponential scalelength.)
This definition ranked with the best in all three categories above. 
The distinction of yielding smallest residuals in TF diagrams is 
shared by essentially any {\it isophotal} linewidths (here, $V_{iso}$ 
and $V_{opt}$). 

It is possible to define 
an optical rotational velocity for TF use which is an identical
match to radio linewidths.  We showed that $\v22$ provides 
such a match for $W_{21cm}^{50}$, the radio linewidth measured at 50\% 
of peak intensity by G97.  The good agreement between optical and 
radio linewidths is understood because a large fraction (50-60\%)
of the \hi surface density is actually contained within the optical radius,
and, in addition, a significant fraction (30\%) of ORCs also reach 
their peak velocity before $r_{opt}$.  

Finally, TF calibrations based solely
on optical linewidths can yield as robust and accurate measures of 
TF distances as analyses based on \hi linewidths, provided the choice 
of optical linewidth is made judiciously. 

Tables 3, 6, and 7, together with the figures in Appendix A will 
be fully released in the AAS CD-ROM Series (vol. 8). 

 
\clearpage

{\noindent {\bf Acknowledgements}}

I would like to thank Michael Strauss for a thorough reading
of an early draft and many valuable suggestions.
Many thanks as well to Adrick Broeils for permission to use data
from his thesis (Fig. 16) and useful discussions on \hi
linewidths and turbulence corrections.  Hans-Walter Rix
offered pertinent comments, especially on \S 5, and Gary
Bernstein clarified some issues regarding $W_{prob}$.
 
I would like to express my gratitude to Jes\'us Gonz\'alez, former
officemate and graduate student at UC Santa Cruz, for patient 
introductions to the art of long-slit reductions.  Sandra Faber
oversaw production of this work at its early stage; she is thanked
warmly for all her support.
 
Vera Rubin kindly supplied paper tracings of ORCs from
photographic plates (\S 4) 
and Alan Dressler also took beautiful data 
for us at Las Campanas Observatory in 1990.  


Much of the data collection and reductions was carried out while 
the author was in graduate school.  Financial support was provided 
by Canadian NSERC and FCAR graduate fellowships as well as by 
Sandra Faber's NSF grant AST 87-02899.


\clearpage

\appendix 
\noindent{\large\bf Appendix A. \ RCs and Models}
\vskip 0.4cm

\ha RCs are presented for 304 Sb-Sc galaxies in
the CF sample.  These are shown on the plane of the sky.  
Multiple observations are not 
overlapped but shown separately to avoid crowding.  A 
total of 376 separate RCs are presented.  For each galaxy, 
two plots are shown with the full two-sided and one-sided curves 
on the left and right panels respectively.  The label above the 
left window for each galaxy shows its name and observing date.
The bottom axis of the right window is given in units of arcseconds
and the top is in kpc (\hub = 70 \hunit).
Error bars have also been omitted to avoid crowding; 
their amplitudes out to the last saved point match roughly the 
size of the data circles.

The center of each RC, about which to fold the curve, was determined 
from Model 2 fits.  On the right panels (for folded RCs), open circles
with a dash at the center represent approaching velocities ($v<0$) while those 
with a cross denote receding velocities ($v>0$).  
The vertical dotted line in the right panels is plotted at 
$r_{disk}$ (=$2.15 h$), the radius of peak rotational velocity 
for a pure exponential disk.  This line is only shown if photometry is 
available in Paper I. 

Models 1 and 2 (see \S 5) are shown by a long-dash line and a continuous 
line respectively.  The URC parameterization of PSS is represented with 
a dotted curve.  The latter could not be plotted if $L_B$ was not listed
in the RC3 or if $r_{opt}$ was not available. 

The full content of Appendix A includes 55 pages of figures.  Only the 
first page is presented here.  The full series will be released in the 
AAS CD-ROM Series (vol. 8). 

\newpage
\noindent{\large\bf Appendix B. \ Model Parameters}
\vskip 0.4cm

Fit parameters for Models 1 and 2 are presented for all 304 CF galaxies
and 958 MAT galaxies.  These are given in Tables 6 and 7 respectively. 
A number of MAT RCs included in the compilation of Persic \& Salucci (1995;
hereafter PS)
were not fitted due to poor signal-to-noise or impossibility to define
a rotation center.  Those are: 
140-g23, 375-g60, 382-g32, 416-g28, 418-g9, 540-g10, and 565-g23. 

The center values ($v_0$ and $r_0$) from Model 1 and 2 for each MAT galaxy
were virtually identical.  For convenience, the centers from Model 2 only
were shown in Table 7. 
As before, the full content of these tables will be released in 
computer-readable files in the AAS CD-ROM Series. 

Our re-derivation of the recession velocity for MAT galaxies 
can be compared with that of PS. 
This comparison is shown in Fig. 18 where we plot the difference between our 
measured $v_0$ and the value determined by PS.  Each galaxy is represented 
by its ESO/LV catalog name. 
The scatter is appreciably larger than what we get by comparing our solutions
between Model 1 and 2, as well as other fitting techniques (we also explored
spline fits to the RC to determine its center).  Different weighting schemes
could explain the discrepancy. 


\notetoeditor{Tables 6 and 7 should be inserted here.}

\newpage
\noindent{\large\bf Appendix C. \ Intensity-Weighted Centroids}
\vskip 0.4cm

To determine the error in the position of the centroids using intensity weighting, 
define $x_i$ and $f_i$ as the position and intensity at each pixel $i$.
The centroiding algorithm used a total of 5 pixels centered about the 
intensity peak, $x_0$, of the emission feature at each row.  The emission-line 
centroid is given by:
\begin{equation}
\left\langle x \right\rangle = \frac{\sum_i (x_i-x_0) f_i }{\sum_{i} f_i}
= \sum_i (x_i-x_0) w_i 
\end{equation}
where we have defined the pixel weights
\begin{equation}
w_i \equiv \frac{f_i}{\sum_i f_i} = \frac{f_i}{F},
\end{equation}
and where $F$ is the total flux under the curve.  The error of the 
centroid is given by

\begin{equation}
\sigma^2\left\langle x \right\rangle = \sum_i (x_i-x_0)^2 \sigma_{w_i}^2
+ \sum_{i,j} (x_i-x_0)(x_j-x_0) \sigma_{w_i w_j}.
\end{equation}

Note that the pixels are oversampled (by a factor two or three)
due to seeing, 
is not correlated since they are sampled independently.  Therefore, the 
covariance term in eq. (4) vanishes and we can write, 
\begin{equation}
\sigma^2\left\langle x \right\rangle = \sum_i {(x_i-x_0)}^2
\left\{ \frac{\sigma_{f_i}^2}{f_i^2} + \frac{\sigma_{F}^2}{F^2} \right\} w_i^2
\end{equation}
where $\sigma_F^2 = \sum \sigma_{f_i}^2$. Both terms $\sigma_{f_i}^2$ and 
$\sigma_F^2$ come from photon statistics and are measured from the variance image. 
The first term inside brackets can be interpreted as the fractional 
error in the intensity at each pixel. The second term gives the signal-to-noise
ratio of the total intensity of the line, {\ie},
\begin{equation}
S/N = \frac{F}{\sigma_{F}}.
\end{equation}


The method of Gaussian fitting yields similar values for the line centers 
(within 0.1 pixels) at S/N $> 20$ but is not as robust as intensity-weighted
centroids in conditions of low S/N.  In an attempt to increase the S/N, we 
also tested ``triple-Gaussian'' fits - those which include the lines of \ha\
and \nii\ simultaneously, but did not achieve appreciable gains.
It should noted that the error term given in Eq. 4 is $\sim$10\% smaller 
than the value obtained via Gaussian fitting. 

The fitting of emission centroids at each row uses a technique of
parabolic-binned interpolation. 
CCD pixel values are an integral of the product of the spectrum and the pixel 
response function.  In bin interpolation, it is the integral of the interpolating 
function rather than the function itself that is constrained to match the pixel 
values.  A parabolic-binned interpolation fits a parabola whose area within each 
pixel (rather than its value at the center of the pixel) matches the pixel values.

Consider a series of CCD pixel values (see Fig. 19) fitted with the parabola

$$ y(x) = a_0 + a_1 x + a_2 x^2 $$

The position of the function maximum ($dy = 0$) is given by
$$\displaylines{\hfill x_{max} = \frac{-a_1}{2a_2} 
  \hspace{.1in} \hbox{and} \hspace{.1in}
  y_{max} = a_0 - \frac{1}{4}\frac{a^2_1}{a_2} \hfill \cr}$$

Let us use for simplicity $x_0=0, x_-=-1,$ and $x_+=+1$.
A standard parabolic fit through each point yields,
\[
\begin{array}{lcl}
 y_0(x_0) &=& a_0             \\
 y_-(x_-) &=& a_0 - a_1 + a_2 \\
 y_+(x_+) &=& a_0 + a_1 + a_2 
\end{array}
\hspace{.2in} \Biggr\} \hspace{.3in}
\begin{array}{rcl}
 a_0 &=&  y_0  \\
 a_1 &=& (y_+ - y_-)/2 \\
 a_2 &=& (y_+ + y_-)/2 - y_0
\end{array}
\]

However, taking the binned nature of the data into account, we must integrate the area 
at all pixels, that is
\[
\begin{array}{lcccl}
y_0 &=& \int_{-1/2}^{1/2} y(x)dx &=& a_0 + \frac{a_2}{12} \\ [9 pt]
y_- &=& \int_{-3/2}^{-1/2} y(x)dx &=& a_0 - a_1 + \frac{13}{12}a_2 \\ [9 pt]
y_+ &=& \int_{1/2}^{3/2} y(x)dx &=& a_0 + a_1 + \frac{13}{12}a_2
\end{array}
\]

Inverting the transformation gives,

\[ 
{\bf \bar a} =
\left\lgroup\matrix{a_0 \cr a_1 \cr a_2 \cr}\right\rgroup =
\left\lgroup\matrix{13/12 y_0 - (y_+ + y_-)/24 \cr
                    (y_+ - y_-)/2 \cr
                    (y_+ + y_-)/2 -y_0 \cr}\right\rgroup
\]

Comparison of this solution for a parabolic-binned interpolation and 
the basic parabolic fit at the pixel centers above shows an offset 
for the terms $a_0$.  The two differ by 
\[
\frac{1}{12}\left(y_0 - \frac{y_+ + y_-}{2}\right).
\]

Our measurements properly account for this difference.


\clearpage
\begin{figure}
\epsscale{1.00}
\caption{
Example of a raw spectrum corrected for geometrical distortion. The galaxy is
UGC~1437. This figure should be compared with Rubin's photographic spectrum for
the same galaxy (NGC 753) shown in Rubin~\etal (1980, Plate 17). The depth of
exposure is comparable (see also Fig. 5, top left panel.}
\end{figure}
\notetoeditor{This figure has no label ``Figure 1'' on it}

\begin{figure}[t]
\epsscale{1.00}
\caption{
Same spectrum as Fig. 1 but with the skylines removed.}
\end{figure}
\notetoeditor{This figure has no label ``Figure 2'' on it}

\begin{figure}[t]
\epsscale{1.00}
\caption{
Spectrum of UGC~1437 with the skylines {\it and} continuum removed. The image is now ready 
for extraction of the rotation curve. Note that the lines \sii\ are barely visible. The continuum 
subtraction is accurate
over most of the chip except for deviations of a percent or so at the position of absorption
lines or where night sky lines and continuum once intersected, as well as long-ward of
6800 \AA\ 
where focus variations become significant.}
\end{figure}
\notetoeditor{This figure has no label ``Figure 3'' on it}


\begin{figure}[t]
\epsscale{1.00}
\caption{
Histogram of the maximal extent of ORCs for our sample.
$r_{max}$ is taken to be the largest radius of detection
on either the approaching or receding side.
The disk exponential scalelength $h$ is measured from
1D bulge-to-disk decompositions (Paper I, Broeils \& Courteau 1997).
The fall-off beyond the optical radius at $3.2h$ is due to the
rapid decline in \ha flux density.}
\end{figure}

\begin{figure}[t]
\epsscale{1.00}
\caption{
Comparisons with the RCs from Rubin and coworkers.  The latter are represented by large
open squares and ours with smaller filled circles (and open circles, filled squares, 
or open triangles for duplicate observations of our own.)  Some eye-ball smoothing was 
applied by Rubin; our data are shown as sampled by the detector (no spatial binning). 
No corrections were applied for inclination or redshift broadening.}
\end{figure}

\begin{figure}[t]
\epsscale{1.00}
\caption{
Comparisons with the RCs MAT.  MAT's data are represented by open 
squares and ours with filled circles.  MAT's data were binned (smoothed)
on a 2\arcsec scale.}
\end{figure}
\clearpage

\setcounter{figure}{6}
\begin{figure}[t]
\epsscale{1.00}
\caption{
 a) Comparisons with the RCs of Vogt (1994).  The point types follow the same
    convention as in Fig. 5 and 6.  The systemic velocities for Vogt's 
    U8488 and U11809 curves were readjusted to better match our RCs.}
\end{figure}

\setcounter{figure}{6}
\begin{figure}[t]
\epsscale{1.00}
\caption{
 b) Comparisons with the RCs of Vogt (1994).}
\end{figure}

\setcounter{figure}{7}
\begin{figure}[t]
\epsscale{1.00}
\caption{
Comparisons with the Fabry-Perot RCs of Amram \etal (1992, 1994).}
\end{figure}

\begin{figure}[t]
\epsscale{1.0}
\caption{
Example of arctan models for the fitting of resolved RCs.  Different 
values of the truncation radius, $r_t$, are shown for six models with 
$v_c = 200$ \kms.  Note that $r_t$ is not strictly equal to the radius 
at turnover.}
\end{figure}

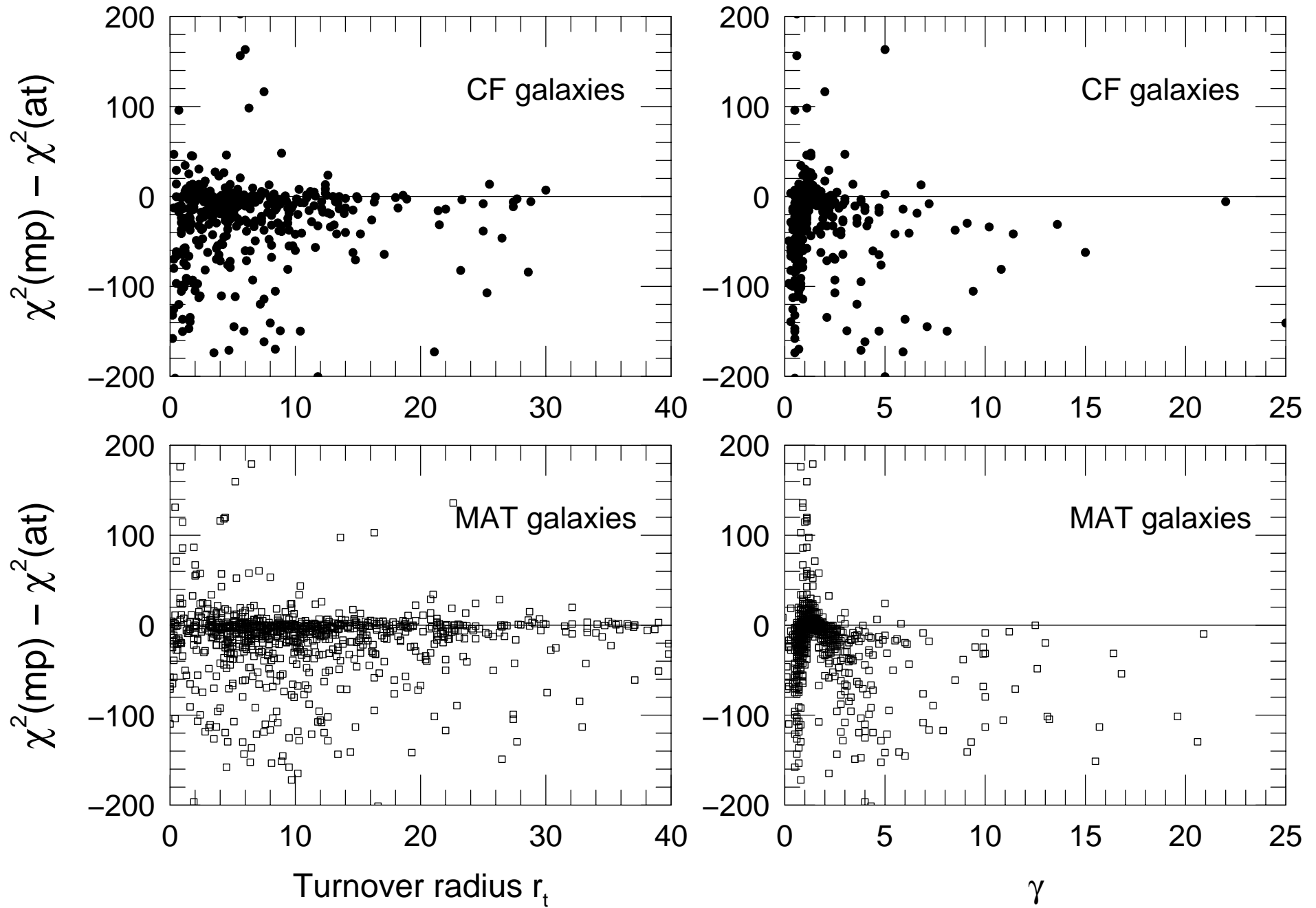
\begin{figure}[t]
\epsscale{1.0}
\caption{
Example of the multi-parameter fitting function for resolved ORCs.
Different values of the shape parameter $\gamma$ are shown for six models
with $v_c$ = 200 \kms\ and $r_t=10$.  For a high-enough $\gamma$, 
$r_t$ does correspond to the exact location of the RC turnover.
Variations with the $\beta$ parameter, though seldom used, 
are also shown as dotted lines.}
\end{figure}

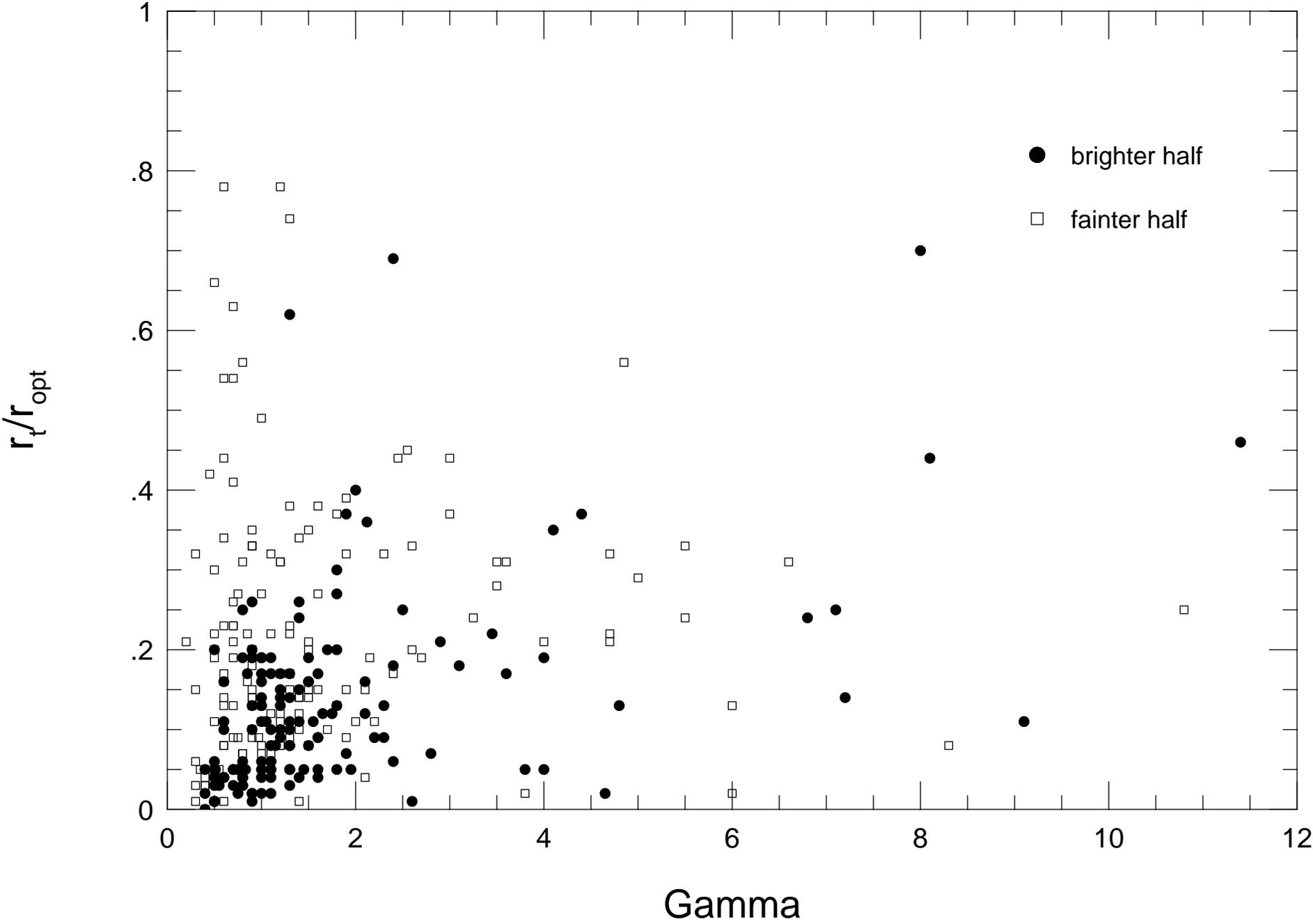
\begin{figure}[t]
\epsscale{1.00}
\caption{
\chisqr\ differences between Model 2 and Model 1 fits for 
CF and MAT galaxies. A preponderance of negative differences 
favors the best fitting function.
The fit parameters $r_t$ and $\gamma$ are from Model 2.}
\end{figure}

\begin{figure}[t]
\epsscale{1.00}
\caption{
 Distribution of $r_t/r_{opt}$ against $\gamma$ for Model 2 with CF galaxies.  
 The parameter coupling is strong for $r_t/r_{opt} \leq .3$ and 
 $\gamma \leq 2.5$.  This explains  the ``hook'' at small $\gamma$'s observed 
 in the right panels of Fig. 11.  The filled circles show the brighter half
 of the CF sample and the open squares represent the fainter half.}
\end{figure}
\clearpage

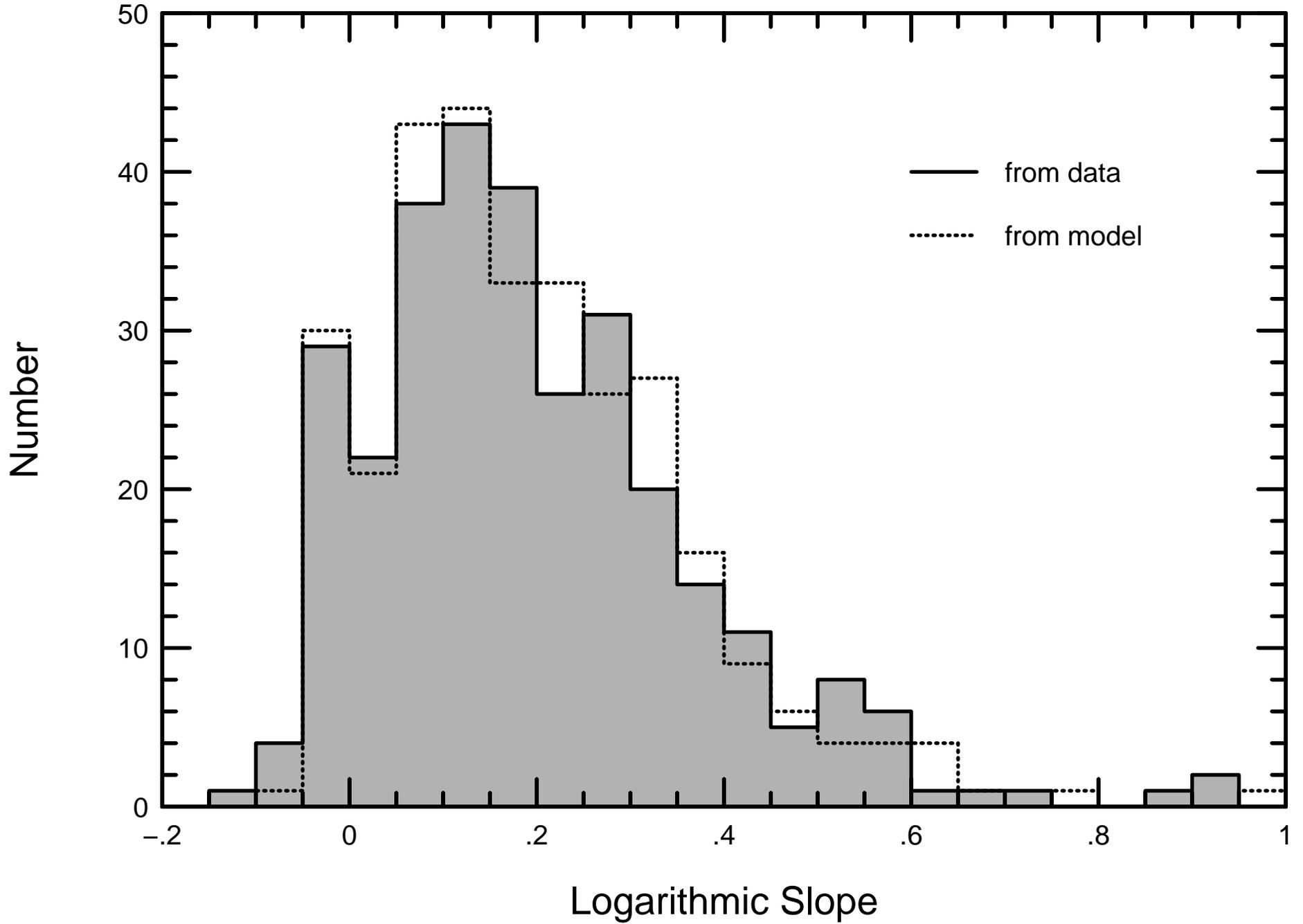
\begin{figure}[t]
\epsscale{1.00}
\caption{
Flux-weighted integrated profile for UGC1437. The various flux levels which define our
measures of rotational linewidths are described in the text.  The vertical arrows
show the profile boundaries inside of which a total area is computed.  The 
fractional intensity levels are determined via parabolic-binned interpolation 
(Appendix C).}
\end{figure}
\notetoeditor{This figure has no label ``Figure 13'' on it}

\begin{figure}[t]
\epsscale{1.00}
\caption{
Histogram of logslopes from data ($ls_r$) and Model 2 ($ls_m$). 
RCs with $ls \leq 0.2$ are deemed convergent.
60\% of all the CF galaxies satisfy that criterion.}
\end{figure}

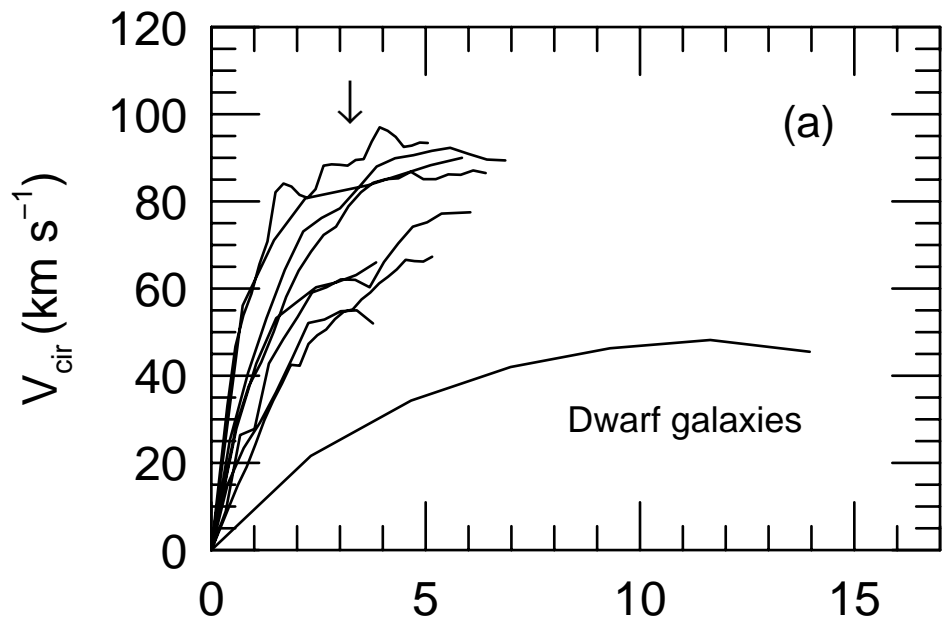
\begin{figure}[t]
\epsscale{1.00}
\caption{
TF calibrations for six definitions of linewidths (described in the text.) 
Open circles correspond to all CF galaxies.  Filled circles represent 
``Quiet Hubble Flow'' galaxies which are less perturbed by streaming 
and local infall motions.
Linewidths are fully corrected, as explained in the text.}
\end{figure}
 
\begin{figure}[t]
\epsscale{1.00}
\caption{
Resolved RCs from \hi  maps of nearby galaxies  (figure adapted with 
permission from Broeils (1992).)  The radius is normalized by the disk 
exponential 
scalelength $h$.  The vertical arrow in each panel indicates the location 
of the optical radius ($r_{opt} \sim 3.2h$).  Most of the intermediate 
and bright galaxies show a velocity peak before or near $r_{opt}$. }
\end{figure}

\begin{figure}[t]
\epsscale{1.00}
\caption{
Comparison of (twice) the maximum rotational speed against
the ratio of rpeak, the radius at maximum velocity, normalized 
by the disk scalelength. 
The open circles data come from CF ORCs (Model 2) which satisfy 
$ls_m \leq 0.2$
and the filled circles are measured from extended \hi RCs from
the compilation of Broeils (1992). 
The horizontal line at 240 \kms\ is a broad demarcation between ``TF'',
fast (HSB galaxies) and slower (LSB and dwarf galaxies) rotators. 
This figure confirms the matching trends between radio and 
ORCs for large galaxies within 4-5 scalelengths.  
The dotted vertical line shows the location of $r_{opt}$. 
The inset shows the histogram for all CF galaxies in this figure.
Radio and optical RCs are shown to peak at roughly the same 
location near the optical radius.  This supports the
good correlation between radio and ORC linewidths for bright 
galaxies (\S 11).}
\end{figure}

\begin{figure}[t]
\epsscale{1.00}
\caption{
Difference in systematic velocities between CF (Model 2)
and the models by Persic \& Salucci (1995) for 958 MAT galaxies.
The points are labelled by their galaxy name.}
\end{figure}

\begin{figure}[t]
\epsscale{1.00}
\caption{
Illustration of parabolic-binned interpolation.  The
pixels centroids are determined by integration of the area 
at all pixels.  This technique yields accurate sub-pixel 
interpolations to estimate fractional flux levels from 
binned data.}
\end{figure}
\clearpage

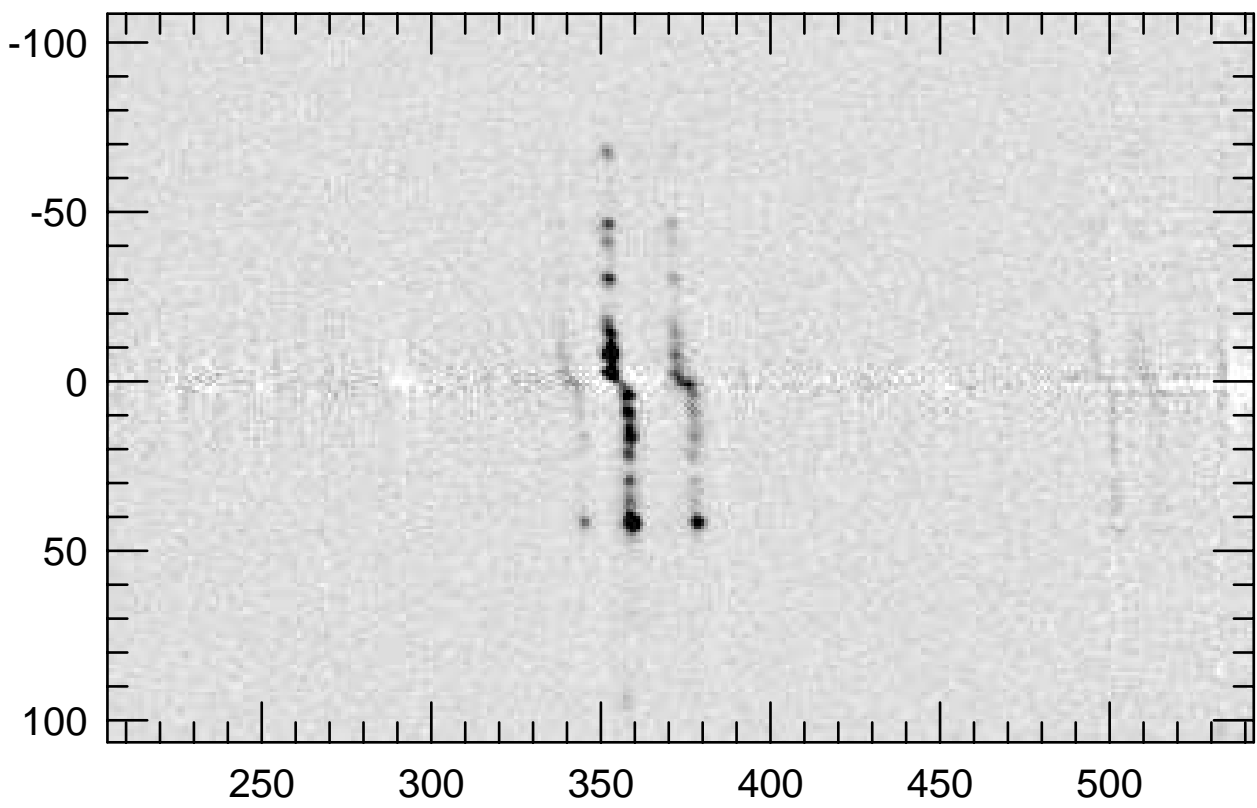
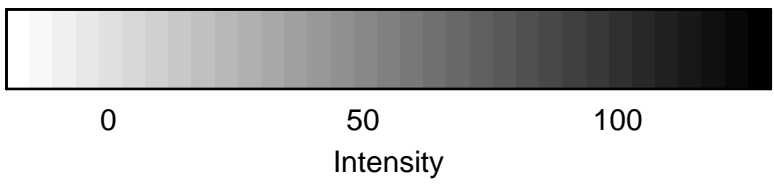
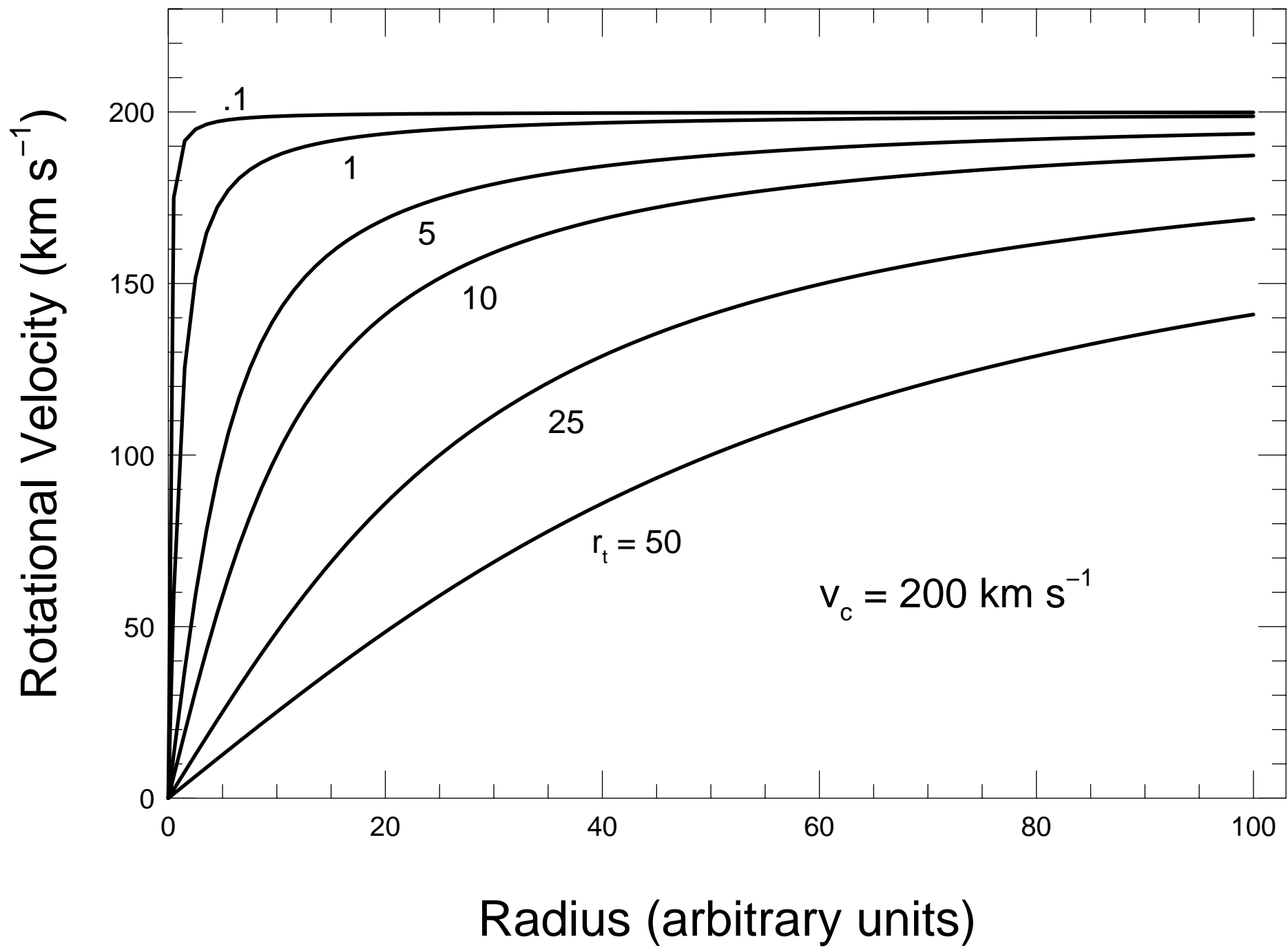
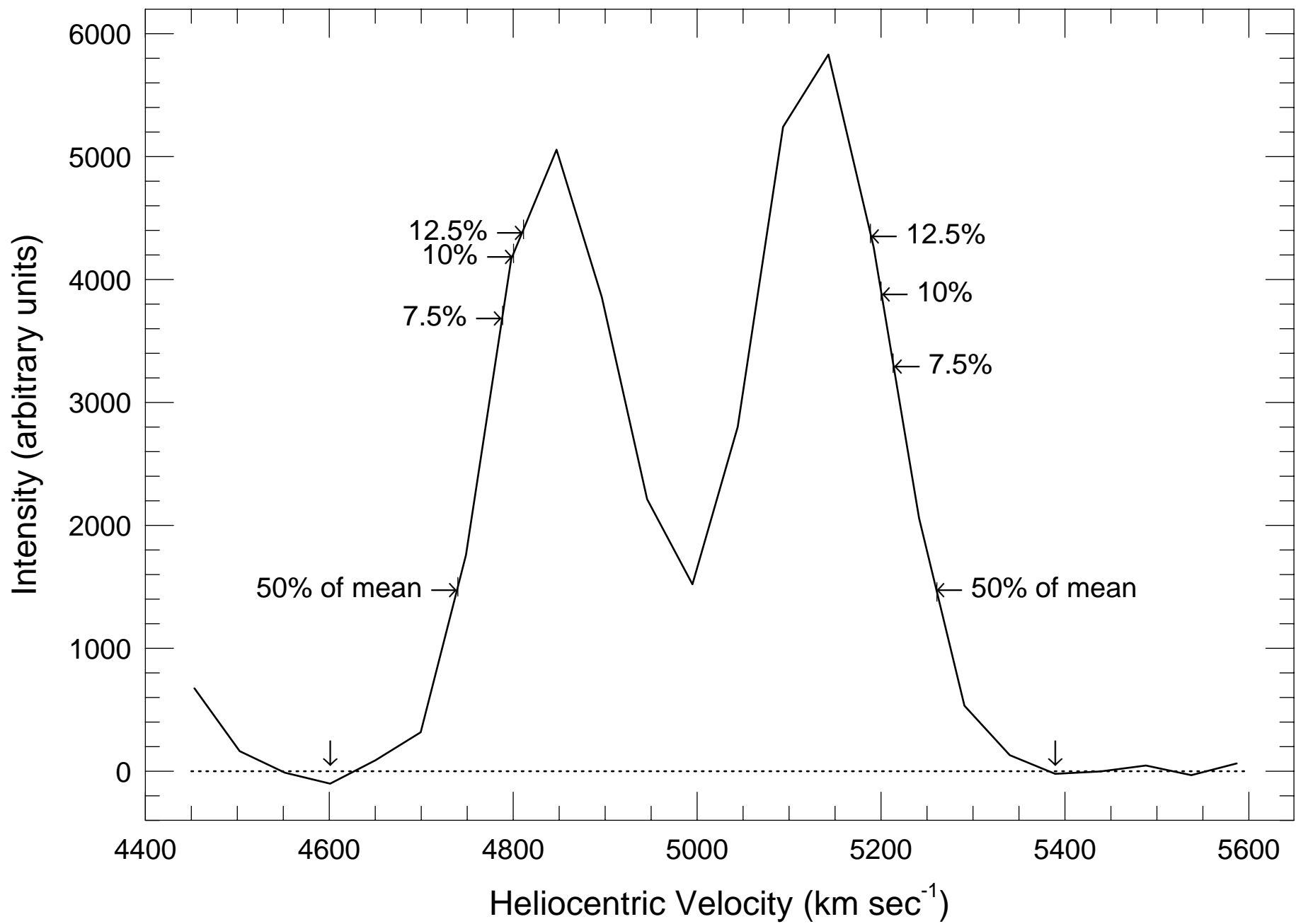
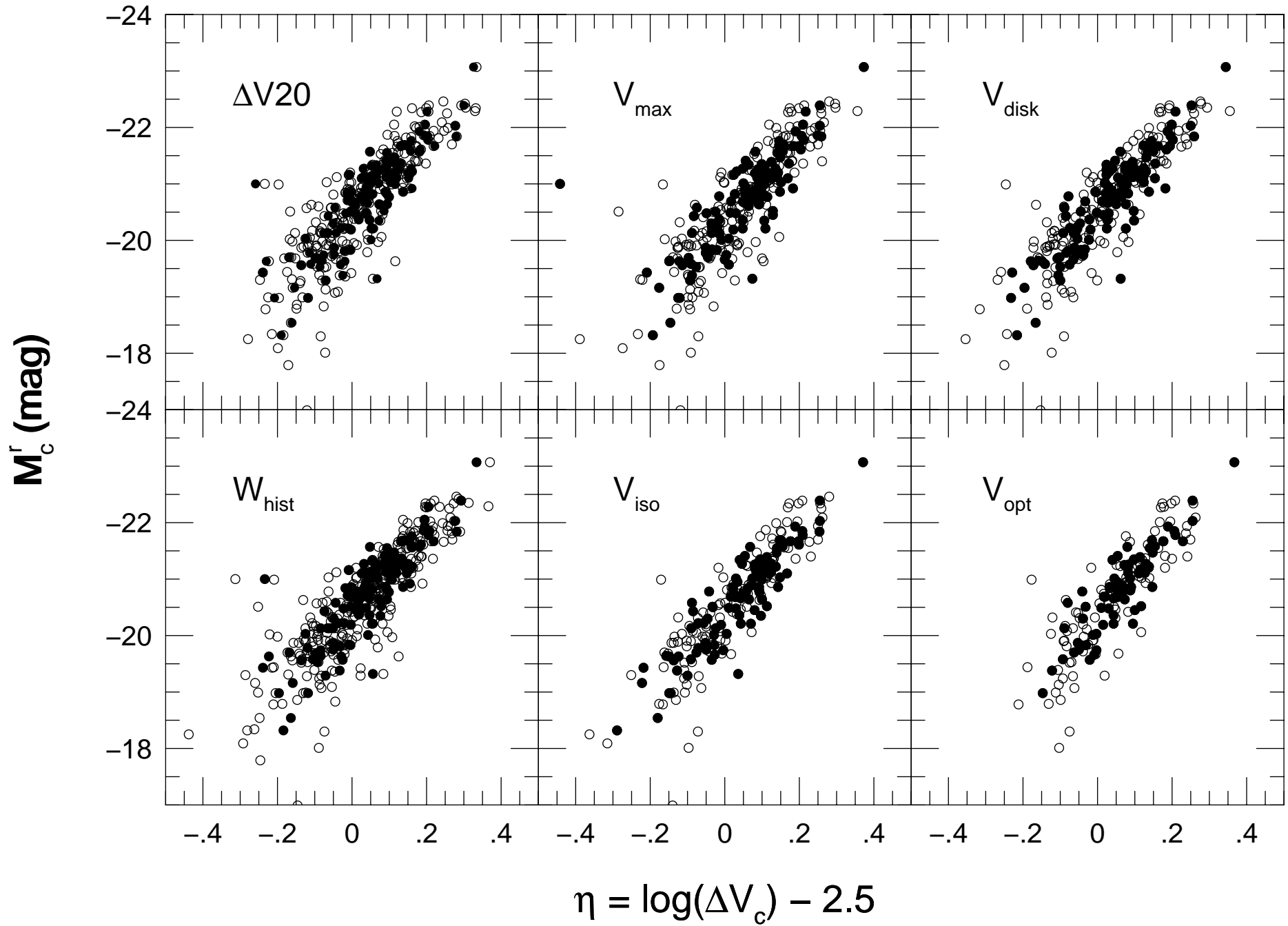
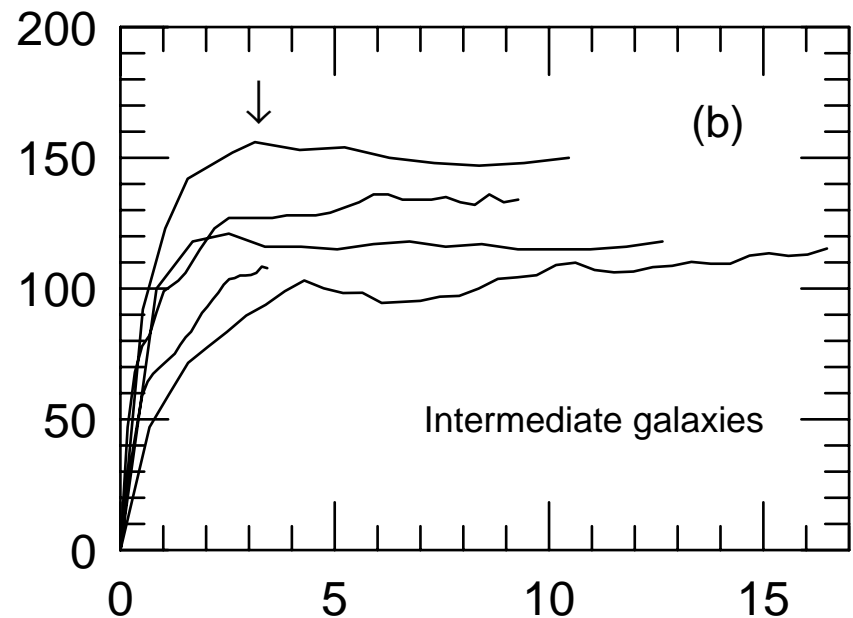
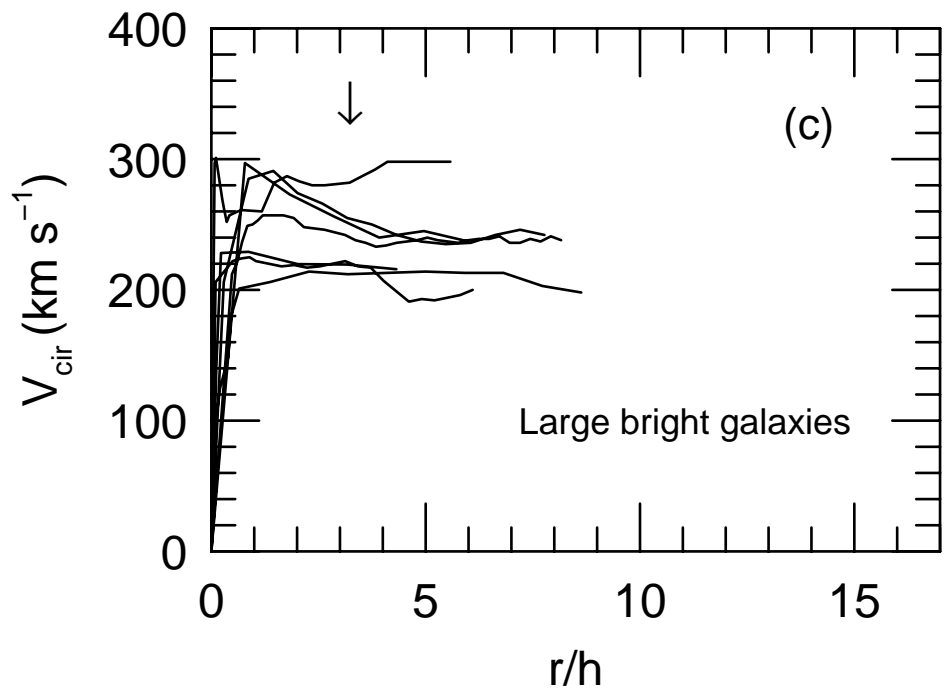
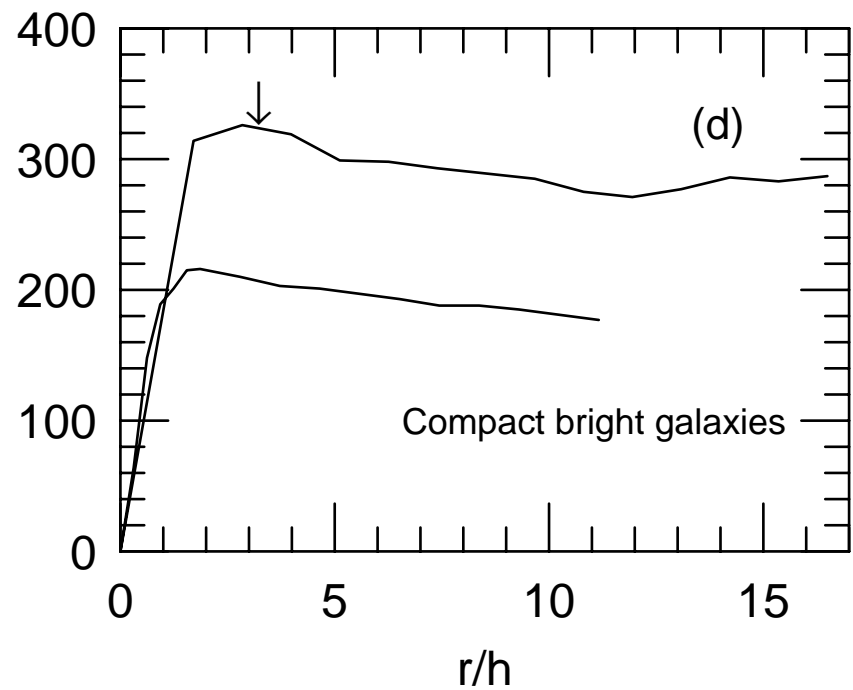
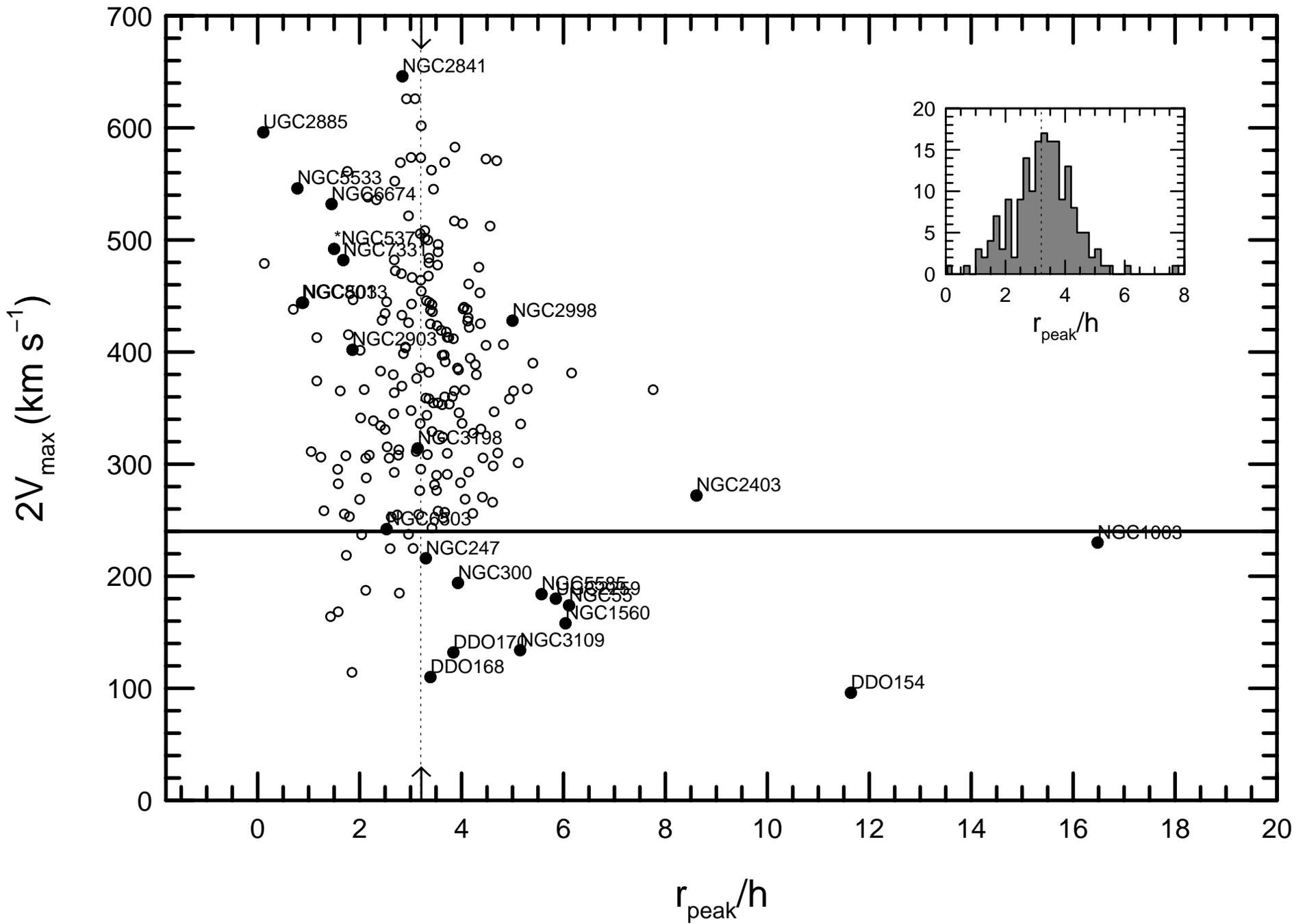
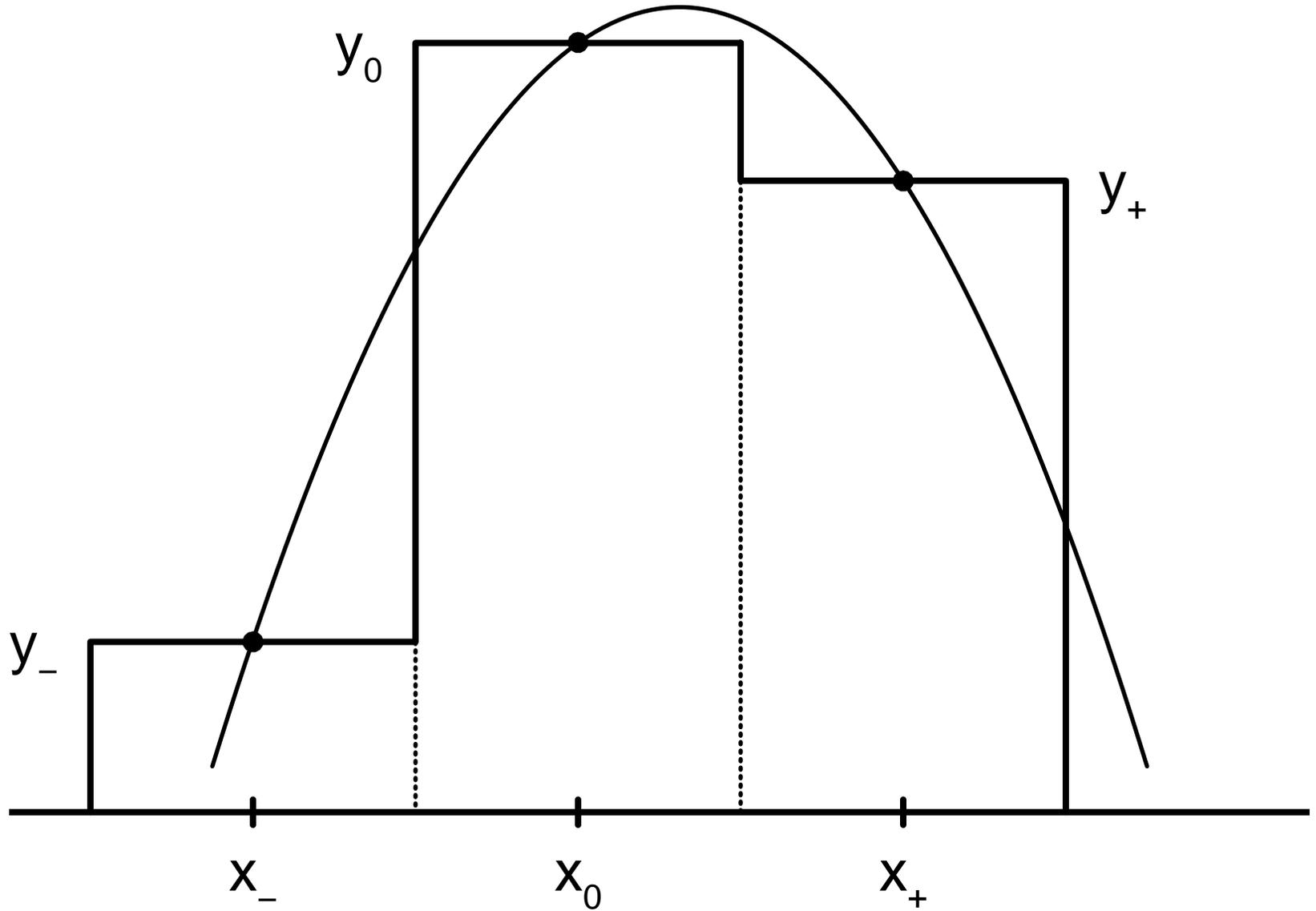
\end{document}